\documentclass[10pt]{article}

\usepackage{amsmath}

\usepackage{array}

\usepackage{appendix}

\usepackage{tocloft}                   

\usepackage{graphicx}

\usepackage{amsfonts}

\usepackage{amssymb}

\usepackage{mathrsfs}

\usepackage{yfonts}

\usepackage{euscript}

\usepackage{centernot}                 

\usepackage{ifsym}                     

\usepackage{lmodern}                   

\usepackage{upgreek}

\usepackage{mathtools}

\usepackage{color}

\usepackage{slantsc}
\usepackage{calligra}

\usepackage{bbold}          

\usepackage[T1]{fontenc}

\usepackage{epsf}

\usepackage{latexsym}

\usepackage{tipa}

\usepackage{makeidx}

\makeindex

\def\b{\begin{equation}}

\def\e{\begin{equation}}

\def\be{\begin{equation}}              

\def\ee{\end{equation}}

\def\beq{\begin{equation}}

\def\eeq{\end{equation}}

\def\bea{\begin{eqnarray}}

\def\eea{\end{eqnarray}}

\def\half{\mbox{$\frac{1}{2}$}}

\def\m{\mbox{ }}

\def\mma {\m , \m \m }

\def\!{\hspace{-1.6667em}}

\def\c{\cite}

\def\l{\label}

\def\r{\ref}

\def\n{\noindent}

\def\u{\underline}

\def\uc{\underbracket}   
\def\w{\widetilde}

\def\bia{\mbox{\boldmath$a$}}

\def\bic{\mbox{\boldmath$c$}}

\def\sbig{\mbox{\scriptsize\boldmath$g$}}

\def\sbiu{\mbox{\scriptsize\boldmath$u$}}

\def\sbip{\mbox{\scriptsize\boldmath$p$}}

\def\bip{\mbox{\boldmath$p$}}

\def\biu{\mbox{\boldmath$u$}}

\def\biC{\mbox{\boldmath$C$}}              

\def\biD{\mbox{\boldmath$D$}}

\def\biG{\mbox{\boldmath $G$}}

\def\biP{\mbox{\boldmath$P$}}

\def\biQ{\mbox{\boldmath$Q$}}

\def\biR{\mbox{\boldmath$R$}}

\def\siA{\mbox{\scriptsize\boldmath$A$}}

\def\sbiM{\mbox{\scriptsize\boldmath$M$}}

\def\bidelta{\mbox{\boldmath$\delta$}}

\def\bslLambda{\mbox{\boldmath$\Lambda$}}           

\def\sbLambda{\mbox{\boldmath\scriptsize$\Lambda$}} 

\def\bslOmega{\mbox{\boldmath$\Omega$}}             

\def\buppsi{\mbox{\boldmath$\uppsi$}}               

\def\biomega{\mbox{\boldmath$\omega$}}              

\def\mA{\mbox{A}}

\def\mF{\mbox{F}}

\def\mI{\mbox{I}}                        

\def\mJ{\mbox{J}}  

\def\mK{\mbox{K}}

\def\mL{\mbox{L}}

\def\mM{\mbox{M}}                        

\def\mN{\mbox{N}}

\def\mS{\mbox{S}}                        

\def\mh{\mbox{h}}

\def\mo{\mbox{o}}

\def\mp{\mbox{p}}

\def\bh{\u{\u{\mbox{h}}}  }            

\def\bE{\mbox{\bf E}}

\def\bP{\mbox{\bf P}}

\def\bQ{\mbox{\bf Q}}

\def\bW{\mbox{\bf W}}

\def\bd{\mbox{\bf d}}

\def\bh{\mbox{\bf h}}

\def\bp{\mbox{\bf p}}

\def\bdelta{\mbox{\boldmath$\delta$}}

\def\buppi{\mbox{\boldmath$\uppi$}}

\def\bupSigma{\mbox{\boldmath$\Sigma$}}                 
\def\sbupSigma{\mbox{\scriptsize\boldmath$\Sigma$}}     

\def\bcalX{\mbox{\boldmath ${\cal X}$}}

\def\fg{\mbox{\tt g}}                          

\def\ft{\mbox{\sffamily t}}                       

\def\fA{\mbox{\sffamily A}}           

\def\bfA{\mbox{\bf\sffamily A}}

\def\bfg{\mbox{\bf\sffamily g}}

\def\fH{\mbox{\sffamily H}}   

\def\fJ{\mbox{\sffamily J}}

\def\fK{\mbox{\sffamily K}}

\def\fP{\mbox{\sffamily P}}

\def\fQ{\mbox{\sffamily Q}}

\def\fR{\mbox{\sffamily R}}

\def\fX{\mbox{\sffamily X}}

\def\cA{{\mathscr A}}

\def\cF{{\mathscr F}}

\def\cG{{\mathscr G}}

\def\cH{{\mathscr H}}

\def\cJ{{\mathscr J}}

\def\cL{{\mathscr L}}

\def\cR{{\mathscr R}}

\def\sa{\mbox{\scriptsize a}}

\def\se{\mbox{\scriptsize e}}

\def\sg{\mbox{\scriptsize g}}

\def\si{\mbox{\scriptsize i}}

\def\sll{\mbox{\scriptsize l}}  

\def\sm{\mbox{\scriptsize m}}

\def\sn{\mbox{\scriptsize n}} 

\def\so{\mbox{\scriptsize o}}

\def\sr{\mbox{\scriptsize r}}

\def\st{\mbox{\scriptsize t}}

\def\sC{\mbox{\scriptsize C}}

\def\sE{\mbox{\scriptsize E}}

\def\sJ{\mbox{\scriptsize J}}

\def\sL{\mbox{\scriptsize L}}

\def\sN{\mbox{\scriptsize N}}

\def\sR{\mbox{\scriptsize R}}

\def\sS{\mbox{\scriptsize S}}

\def\sT{\mbox{\scriptsize T}}

\def\sfA{\mbox{\sffamily{\scriptsize A}}}     

\def\sfK{\mbox{\sffamily{\scriptsize K}}}      

\def\sfN{\mbox{\sffamily{\scriptsize N}}}      

\def\sfX{\mbox{\sffamily{\scriptsize X}}}      

\def\sbd{\mbox{{\bf \scriptsize d}}}

\def\sbfC{\mbox{\bf \scriptsize\sffamily C}}

\def\sbfg{\mbox{\bf \scriptsize\sffamily g}}

\def\bfg{\mbox{\bf \sffamily g}}

\def\sbfP{\mbox{\bf \scriptsize\sffamily P}}

\def\sbfQ{\mbox{\bf \scriptsize\sffamily Q}}

\def\sbcC{\mbox{\boldmath \scriptsize ${\cal C}$}}

\def\sbcF{\mbox{\boldmath \scriptsize ${\cal F}$}}

\def\sbcG{\mbox{\boldmath \scriptsize ${\cal G}$}}

\def\sbcL{\mbox{\boldmath \scriptsize ${\cal L}$}}

\def\sbcS{\mbox{\boldmath \scriptsize ${\cal S}$}}

\def\tcF{\mbox{\tiny ${\cal F}$}}

\def\bfg{\mbox{{\bf \sffamily g}}}                                    

\def\bfQ{\mbox{{\bf \sffamily Q}}}                                    

\def\bfP{\mbox{{\bf \sffamily P}}}                                    

\def\btcP{\mbox{\boldmath\tiny${\cal P}$}}                            

\def\bfu{\mbox{{\bf \sffamily u}}}

\def\btcF{\mbox{{\boldmath \tiny${\cal F}$}}}                                     

\def\bscC{\mbox{{\boldmath \scriptsize${\cal C}$}}}                               

\def\bscF{\mbox{{\boldmath \scriptsize${\cal F}$}}}                               

\def\bscQ{\mbox{{\boldmath \scriptsize${\cal Q}$}}}  

\def\bscR{\mbox{{\boldmath \scriptsize${\cal R}$}}}  

\def\bscM{\mbox{{\boldmath \scriptsize${\cal M}$}}}  

\def\bscK{\mbox{{\boldmath \scriptsize${\cal K}$}}}                              

\def\bscP{\mbox{\boldmath\scriptsize${\cal P}$}}                                 

\def\bscS{\mbox{\boldmath \scriptsize${\cal S}$}}                                

\def\btcF{\mbox{{\boldmath \tiny${\cal F}$}}}                                    %

\def\bfc{\mbox{{\bf \sffamily c}}}                                               

\def\Thomas{\,\,\mbox{\textcircled{$\rightarrow$}}\,\,}

\def\TwoWay{\,\,\mbox{\textcircled{$\leftrightarrow$}}\,\,}

\def\sumi2{\sum\mbox{}_{\mbox{}_{\mbox{\scriptsize $i$=1}}}^2}

\def\sumi3{\sum\mbox{}_{\mbox{}_{\mbox{\scriptsize $i$=1}}}^3}

\def\sumABcycles3{\sum\mbox{}_{\mbox{}_{\mbox{\scriptsize cycles $A,B$=1}}}^{3}}

\def\sumCDcycles3{\sum\mbox{}_{\mbox{}_{\mbox{\scriptsize cycles $C,D$=1}}}^{3}}

\def\sumj3{\sum\mbox{}_{\mbox{}_{\mbox{\scriptsize $j$=1}}}^3}

\def\sumk3{\sum\mbox{}_{\mbox{}_{\mbox{\scriptsize $k$=1}}}^3}

\def\prodiA1{\prod\mbox{}_{\mbox{}_{\mbox{\scriptsize $i$=1}}}^{A - 1}}

\def\d{\textrm{d}}                                                  

\def\pa{\partial}                                                   

\def\bpa{\mbox{\boldmath$\partial$}}                                

\def\ordial{\bd\hspace{-0.088in}\pa}                                

\def\partional{\bdelta\hspace{-0.08in}\pa}                          

\def\sordial{{\sbd\hspace{-0.075in}\mbox{\scriptsize$\pa$}}}        

\def\lordial{\mbox{\large $\bd\hspace{-0.093in}\pa$}}               

\def\es{\m = \m}

\def\:={\m := \m}

\def\=:{\m =: \m}

\def\FrT{\mathfrak{T}}                                         

\def\lFrg{\mbox{\Large$\mathfrak{g}$}}                         
\def\nFrg{\mbox{\large$\mathfrak{g}$}}                         
\def\Frg{\mbox{\normalsize $\mathfrak{g}$}}                    

\def\bFrF{\mbox{\boldmath$\mathfrak{F}$}}

\def\sFrT{\mbox{\scriptsize$\mathfrak{T}$}}                    

\def\bFrP{\mbox{\Large $\mathfrak{p}$}}                        
\def\Hilb{\mbox{{\boldmath$\mathfrak{H}$}ilb}}                 

\def\scC{\mbox{\scriptsize ${\cal C}$}}                    

\def\scD{\mbox{\scriptsize ${\cal D}$}}                    

\def\scE{\mbox{\scriptsize ${\cal E}$}}                    

\def\scF{\mbox{\scriptsize ${\cal F}$}}

\def\btcF{\mbox{\bf\tiny ${\cal F}$}}

\def\scG{\mbox{\scriptsize ${\cal G}$}}                    

\def\scH{\mbox{\scriptsize ${\cal H}$}}                    

\def\scL{\mbox{\scriptsize ${\cal L}$}}                    

\def\scM{\mbox{\scriptsize ${\cal M}$}}                    

\def\scQ{\mbox{\scriptsize ${\cal Q}$}}                    

\def\scS{\mbox{\scriptsize ${\cal S}$}}                    

\def\bLin{\sbcL\mbox{\bf in}} 

\def\Flin{\scF\mbox{lin}}                                  

\def\bFlin{\sbcF\mbox{\bf lin}}

\def\Quad{\scQ\mbox{uad}}                                  

\def\Chronos{\scC\mbox{hronos}}                            

\def\bGauge{\sbcG\mbox{\bf auge}}

\def\Shuffle{\scS\mbox{huffle}}                            

\def\bShuffle{\sbcS\mbox{\bf huffle}} 

\def\FrQ{\mbox{\Large $\mathfrak{q}$}}                               
\def\sFrQ{\mbox{\large $\mathfrak{q}$}}                              
\def\bFrC{\mbox{\boldmath$\mathfrak{C}$}}                            
\def\Phase{\mbox{{\boldmath$\mathfrak{P}$}hase}}                     

\def\bFrR{\mbox{\boldmath$\mathfrak{R}$}}                            
\def\Rig-Phase{\bFrR\mbox{ig-}\Phase}                                
                                                                													   
\def\bFrR{\mbox{\boldmath$\mathfrak{R}$}}                            

\def\bFrR{\mbox{\boldmath$\mathfrak{R}$}}                            

\def\1mat{\u{\u{1}}}                                                 

\def\Positive-Modespace{\mbox{{\boldmath$\mathfrak{M}$}odespace$^+$}}

\def\POSITIVE-MODESPACE{\mbox{{\boldmath$\mathfrak{M}$}ODESPACE$^+$}}
                                                                                                                             														
\def\bFrG{\mbox{ $\mathfrak{G}$}}                                    %

\def\lE{\mbox{\Large E}}

\def\Kin-Hilb{\mbox{{\boldmath$\mathfrak{K}$}in-\Hilb}}                     

\def\Mid-Hilb{\mbox{{\boldmath$\mathfrak{M}$}id-\Hilb}}                     

\def\Dyn-Hilb{\mbox{{\boldmath$\mathfrak{D}$}yn-\Hilb}}                     

\def\5Star{\mbox{\Large$\star$}}              

\def\K{Kucha\v{r} }

\begin{document}

\begin{center}

\Huge{\bf A LOCAL RESOLUTION OF}

\vspace{.1in}

\normalsize

\Huge{\bf THE PROBLEM OF TIME}

\vspace{.15in}

\Large{\bf VII. Constraint Closure} 

\vspace{.15in}

{\large \bf E.  Anderson}$^1$

\vspace{.15in}

{\large \it based on calculations done at Universit\'{e} Paris VII} 

\end{center}

\begin{abstract}

We now set up Constraint Closure in a manner consistent with Temporal and Configurational Relationalism. 
This requires modifying the Dirac Algorithm -- which addresses the Constraint Closure Problem facet of the Problem of Time piecemeal -- 
to the TRi-Dirac Algorithm.
This is a member of the wider class of Dirac-type algorithms that enjoys the property of being Temporal Relationalism implementing (TRi).  
Constraint algebraic structures ensue. 
We include examples of types of constraint, outcomes of the Dirac Algorithm and different kinds of Constraint Closure Problems.  
Enough new Principles of Dynamics is required to support this venture that an Appendix on it is provided: 
differential Hamiltonians, anti-Routhians, and the brackets, state spaces and morphisms corresponding to these.  

\end{abstract}

$^1$ dr.e.anderson.maths.physics *at* protonmail.com

\section{Introduction}\label{QBI-Underpin} 

This is our seventh Article \cite{I, II, III, IV, V, VI} on the Problem of Time 
\cite{Battelle, DeWitt67, Dirac, K81, K91, K92, I93, K99, APoT, FileR, APoT2, AObs, APoT3, ALett, ABook, A-CBI} and its underlying Background Independence. 
Herein, we extend Articles V and VI's consistent unified treatment of Article  I's  Temporal        Relationalism  
                                                                   and Article II's Configurational Relationalism  
																   to consistently deal with Constraint Closure as well.

\m 

\n Having introduced Field Theory since Article III's finite account, the Finite--Field portmanteau notation for constraints is 
\be 
\sbcC  \:=  \cF\lfloor \bfQ, \bfP \rfloor \m : \m  \mF(\biQ, \biP) \m            \mbox{ (finite) } \m \mbox{  and } \m  
                                                   {\cal F}(\u{x}; \bQ, \bP] \m  \mbox{ (field) } .  
\ee
The combination of working in Hamiltonian variables $(\bfQ, \bfP)$ 
and making use of the classical Poisson brackets -- extended to portmanteau notation in Sec 2 --
turns out to allow for a {\sl systematic} treatment of constraints: the Dirac Algorithm \cite{Dirac}.  
We already provided this in Article III; we now generalize this to Dirac-type Algorithms in Sec 3; 
it is the TRi such that we adopt for Problem of Time facet consistency.
We phrase this approach as starting with a {\sl trial action} ${\cal S}^{\st\sr\si\sa\sll}$ producing {\it trial constraints}. 
In cases in which Constraint Closure is completed, `trial' names and labels are promoted to `CC' ones, standing for `Closure completed' 
as well as for the Constraint Closure aspect and facet name.  
Types of constraint are discussed in Sec 4, and constraint algebraic structures in Sec 5. 

\m 

\n {\bf Constraint Closure} - the third Background Independence aspect - is itself considered in Sec 6.  
Complications and impasses with this are the corresponding third facet of the Problem of Time: the Constraint Closure Problem.  

\m 

\n{\bf Functional Evolution Problem} was the facet name used by \K and Isham \cite{K92, I93} in the quantum-level field-theoretic setting.
Some parts of this problem, however, already occur in finite examples, for which partial rather than functional derivatives are involved.   

\m 

\n{\bf Partional Evolution Problem} is thus a more theory-independent portmanteau name for this facet.  

\m 

\n{\bf Constraint Closure Problem} is a more general facet name through its additionally covering the classical version of the problem to some extent.    

\m 

\n As useful recollection and for reference, some parts of Problem of Time facet composition 
between Constraint Closure and Temporal and Configurational Relationalism already given in Article III are as follows.

\m 

\n The split Constraint Provider input has, 
on the   one hand,        Temporal Relationalism provides a constraint $\Chronos$ that is quadratic and so is also denoted by $\Quad$.    
On the other hand, Configurational Relationalism provides candidate   $\bShuffle$ constraints that are linear and so are also denoted by $\bLin$.  
One is then to use the Dirac Algorithm on this combined incipient set of constraints 
to see whether Constraint Closure is met or the Constraint Closure Problem arises. 
This split induces a further split consideration of Constraint Closure: whether each of $\Chronos$ and $\bShuffle$ are self first-class, 
                                                                             and whether they are mutually first-class.
The self and mutual behaviour of $\bShuffle$ determines whether Configurational Relationalism has succeeded.
Let us also point to the useful `end summary road map' Fig \ref{TR-CR-CC-Summary} as regards keeping track of how these various facet interferences fit together.

\m 

\n Enough new Principles of Dynamics is required to support this venture to merit an Appendix.  
This covers differential Hamiltonians, ordial \c{VI} differential almost-Hamiltonians, anti-dRouthians, 
Peierls brackets \c{Peierls} and mixed Poisson--Peierls brackets, and the corresponding state spaces and morphisms.

\section{Poisson brackets and phase space}

\n{\bf Structure 1} As a first instance of Equipping with Brackets,   
consider the joint space of the $\bfQ$ and $\bfP$ alongside the classical Poisson brackets
\beq
\mbox{\bf \{}    F    \mbox{\bf ,} \, G  \mbox{\bf \}}  \:=  \int_{\sN\so\sS}\d \mN\mo\mS
\left\{
\frac{\partional F}{\partional \bfQ} \cdot  \frac{\partional G}{\partional \bfP}  -  
\frac{\partional F}{\partional \bfP} \cdot  \frac{\partional G}{\partional \bfQ} 
\right\}                                                                                           \m , 
\label{PB-Port}
\eeq
i.e.\ the portmanteau of (III.1) for finite theories and 
\be 
\mbox{\bf \{}    F    \mbox{\bf ,} \, G  \mbox{\bf \}}  \:=  \int_{\sbupSigma}\d \bupSigma
\left\{
\frac{\updelta F}{\updelta \bQ} \cdot  \frac{\updelta G}{\delta \bP}  -  
\frac{\updelta F}{\updelta \bP} \cdot  \frac{\updelta G}{\delta \bQ} 
\right\} 
\label{TRi-Field-PB}
\ee 
for Field Theories. 

\m 

\n The fundamental Poisson bracket is 
\beq 
\mbox{\bf \{}    \u{\bfQ}    \mbox{\bf ,} \,   \u{\bfP} \mbox{\bf \}}  \es  \u{\u{\bidelta}}
\eeq
for $\bidelta$ the portmanteau of the finite Kronecker $\delta$ 
and the product of a species-wise such with a field-theoretic Dirac $\updelta^{(d)}(\underline{x} - \underline{x}^{\prime})$.  
This bracket being established for all the $\bfQ$ and $\bfP$ means that brackets of all once-differentiable quantities
$\cF\lfloor \bfQ$, $\bfP \rfloor$ are established as well.

\m  

\n{\bf Remark 1} The entries into each slot of the Poisson brackets could also be functionals $\cF$, $\cG$ rather than just functions $F$, $G$.    

\m

\n{\bf Structure 1} In terms of Poisson brackets, the equations of motion are
\beq
\dot{\bfQ} = \mbox{\bf \{} \bfQ \mbox{\bf ,} \, \cH \mbox{\bf \}}                               \mma  
\dot{\bfP} = \mbox{\bf \{} \bfP \mbox{\bf ,} \, \cH \mbox{\bf \}}                               \m .   \l{PEOM}
\eeq
\n{\bf Structure 2} Thus equipped, this joint space is known as {\it phase space}, $\Phase$. 

\m 

\n{\bf Remark 2}  Our first preoccupation is establishing which structures are already TRi, and which need to be supplanted by TRi counterparts.
The Poisson bracket is already-TRi. 
(\r{PEOM}) becomes 
\beq
\ordial{\bfQ} = \mbox{\bf \{} \bfQ \mbox{\bf ,} \, \cH \mbox{\bf \}}                               \mma  
\ordial{\bfP} = \mbox{\bf \{} \bfP \mbox{\bf ,} \, \cH \mbox{\bf \}}                               \m . 
\eeq
$\Phase$ is already-TRi, since all of $\biQ$, $\biP$ and the Poisson bracket are.

\m 

\n{\bf Structure 3} The Liouville 1-form 
\be
\biP  \cdot  \ordial \biQ
\ee 
and the symplectic 2-form 
\be 
\ordial \biP  \wedge \ordial \biQ 
\ee
-- each further motivated in Appendix \r{Mor} --
are relatively rare examples of already-TRi objects of nonzero weight:  change 1- and 2-forms respectively.
See the Appendix for some of their further significance.  

\m 

\n As the inverse of the previous, the Poisson tensor $\biC$ is recast as a change 2-tensor $d^{-2} \biD$.  

\m 

\n{\bf Structure 4} Temporal Relationalism also requires use of specifically time-independent canonical transformations $Can$ 
rather than the more general $Can_t$ (see Appendix \r{Mor}) in the $d$-Hamiltonian formulation.  

\vspace{10in}

\section{Dirac-type Algorithm}\label{Dirac-Proc-Purpose}

\subsection{TRi appending via cyclic differentials}

\n{\bf Remark 1} One part of handling constraints is to additively append them to a bare Hamiltonian-type object.
For this to attain TRi, the bare object is to be a differential Hamiltonian 
\be 
\ordial \cH   \m .
\ee 
and the appending is to be done not with cyclic differentials in place of Dirac's Lagrange multipliers.  

\m 

\n{\bf Structure 1} The TRi Dirac-type Algorithm requires declaring differential-almost-Hamiltonian variables: $(\bfQ, \bfP)$ in the part-physical sector 
                                                                                   and $\ordial \bfA$ in the purely auxiliary sector.

\m 

\n{\bf Definition 1} The {\it arbitrary-primary differential-almost-Hamiltonian} (alias {\it Dirac-type `starred' differential-almost-Hamiltonian}) is
\be 
\ordial \cA_{\siA\btcP} :=  \ordial \cA^*  
                        :=  \ordial \cH + \ordial \uc{\bfA} \cdot \uc{\bscP}  \m . 
\ee 
I.e.\, the result of taking a bare differential Hamiltonian                         $\ordial \fH$
and additively appending to it a theory's formalism's primary constraints           $\bscP$ 
using arbitrary functions of $(\bfQ, \bfP)$ now represented as cyclic differentials $\d \bfA(\bfQ, \bfP)$.   

\m 

\n{\bf Definition 2} The {\it unknown-primary differential-almost-Hamiltonian} (alias {\it Dirac-type total differential-almost-}

\n{\it Hamiltonian})\footnote{The hyphening used is intended to clarify that these are {\sl in no way implied} to be total differentials.} is 
\be 
\ordial \cA_{\sbiu\btcP} :=  \ordial \cA_{\sT}  :=  \ordial \cH + \ordial \uc{\bfu} \cdot \uc{\bscP}      \m .
\ee 
I.e.\ the result of additively appending the same but now using {\it unknown cyclic differentials} $\ordial \bfu(\bfP, \bfQ)$.  

\m  

\n{\bf Remark 3} As an indication of how the preceding definition is used, 
\beq
0 \m \approx \m  \ordial \bscP    
        \es      \mbox{\bf\{} \bscP \mbox{\bf,} \, \ordial \cA_{\sbiu\btcP} \mbox{\bf\}}
        \es      \mbox{\bf\{} \bscP \mbox{\bf,} \, \ordial \cH \mbox{\bf\}} +  \mbox{\bf \{} \bscP  \mbox{\bf ,} \,  \uc{\bscP} \mbox{\bf \}} \cdot \ordial \uc{\bfu}   \m .
\label{de-for-u}
\eeq
The first step is dictated by consistency, the second by the differential-almost-Hamilton's equations (Appendix A.1), and the third by definition and linearity.

\m

\n{\bf Remark 4} For Field Theories with fields $\uppsi$ and allowing for the spatial metric $\bh$ to be among these, 
we also now require {{\it TRi-smearing} for our constraints:  
\beq 
(    \u{\bscC}    \, | \,    \pa \u{\bW}    )    \:=    \int d^3 x \, \u{\bscC}(\underline{x}; \bh, \buppsi, \bp, \buppi^{\uppsi}] \, \pa \bW(\underline{x}) \m .
\eeq
Such expressions are then inserted inside the classical brackets.  
 
\m  

\n{\bf Remark 5} The brackets in use are now the physical Poisson brackets sector of a mixed Poisson--Peierls bracket (Appendix \ref{dA-Dir}) 
as is appropriate to the corresponding unreduced ordial-almost-phase space $\ordial\mA$-$\Phase$ rather than phase space $\Phase$.  

\m 

\n The complementary Peierls sector (Appendix \cite{Peierls}) is purely-auxiliary, and so is just `unphysical fluff'.

\m 

\n This split is guaranteed by constraints being of the form $\scC(\bfQ, \bfP \mbox{ alone})$, because 
\beq
\mbox{passage from $\ordial \bfQ$ to $\bfP$ absorbs all the $\ordial \fg$}                   \m .
\label{key}
\eeq
This is the {\sl major} trick which can be performed with Appendix A.2's ($\ordial$-)anti-Routhian.  

\m 

\n{\bf Remark 6} The resultant equation (\ref{de-for-u}) is an explicit equation, 
with the cyclic differential auxiliaries $\ordial \biu$ playing the role of unknowns.  

\m 

\n{\bf Remark 7} Since only the Poisson bracket part acting on the constraints, 
the definitions of first- and second-class remain unaffected, as are the Dirac bracket and the extension procedure.

\m 

\n{\bf Remark 8} See Fig \ref{TRiPoD-Squares} for some context.
Phase space $\Phase$ is now replaced by  A-$\Phase$ and $\ordial$A-$\Phase$; 
these are all types of bundle twice over: cotangent bundles {\sl and} $\lFrg$-bundles.  
Locally (in configuration space) product spaces will do.  
%
{            \begin{figure}[!ht]
\centering
\includegraphics[width=0.7\textwidth]{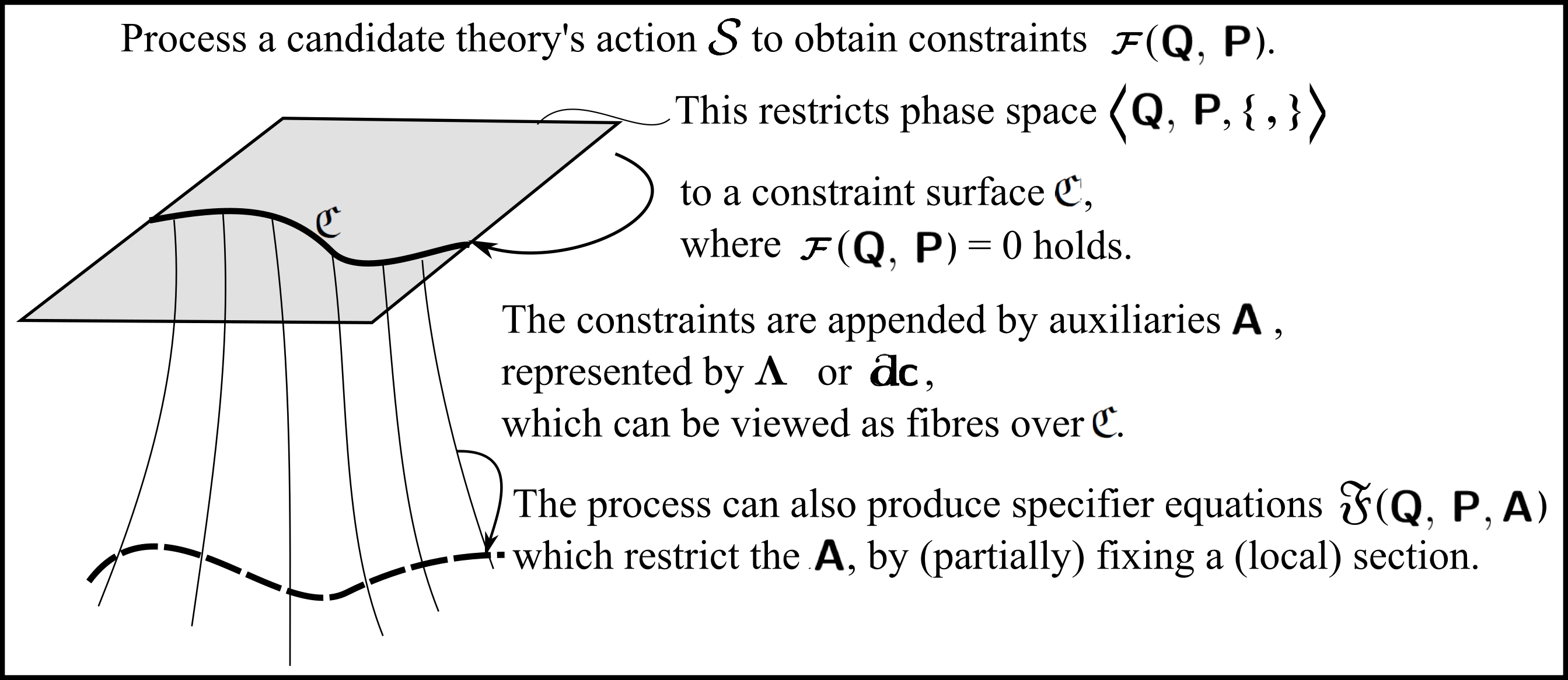}
\caption[Text der im Bilderverzeichnis auftaucht]{ \footnotesize{Geometrical sketch of the outcome of Dirac-type procedures for classical Constraint Closure.  
We omit a loop back to readjusting some of the phase space structures -- $\sbfQ$, $\sbfP$ and classical bracket -- 
in the event of second-class constraints arising.     
While the                  Dirac Algorithm's  Lagrange multipliers  $\sbLambda$      can be incorporated by extending the $\sbfQ$, 
      the TRi-Dirac-type Algorithm's cyclic differentials           $\sordial \sbfC$ are more heterogeneous.
In any case, envisaging the auxiliary variables used in Constraint Appending as fibres 
keeps them separated out from the phase space which contains actual physical information.  
[See Part III of \cite{ABook} for comparison with the simpler assessment of Constraint Closure at the quantum level.]  }        }
\label{DTA}
\end{figure}  }

\subsection{TRi Dirac Little Algorithm}\label{TRi-DA}

Dirac's Algorithm is moreover not TRi because the appending Lagrange multipliers break this. 
It is, instead, supplanted by the TRi-Dirac Algorithm in which appending is performed by cyclic differentials.  

\m 

\n As regards the six cases this is capable of producing at each step (c.f.\ Article III), in any combination, equation types 0), 1) and 3) are as before.
On the other hand, equation types 2) and 3) are now phrased in terms of cyclic differentials.

\m 

\n{\bf Definition 1} {\bf The TRi-Dirac Little Algorithm} \cite{Dirac} consists of evaluating classical brackets between a given set of constraints 
so as to determine whether these are consistent and complete.  

\m 

\n At this level, four types of equation can emerge.

\m 

\n{\bf Type 0)} {\bf Inconsistencies}.

\m 

\n{\bf Type 1)} {\bf Mere identities}.

\m 

\n{\bf Type 2)} {\bf Further secondary constraints}, i.e.\ equations independent of the cyclic differential unknowns. 

\m 

\n{\bf Type 3)} {\bf Specifier equations}, i.e.\ relations amongst some of the appending cyclic differential functions themselves. 
%

\m 

\n{\bf Remark 1} The possibility of specifier equations stems from the TRi Dirac Algorithm involving an {\it appending procedure} involving such auxiliaries. 
As already mentioned in Article III, the Dirac Algorithm itself involves restrictions on Lagrange multiplier auxiliaries. 

\m 

\n{\bf Remark 2} Dirac-type algorithms moreover generalize the type of appending auxiliaries; 
The TRi-Dirac Algorithm subcase involves, concretely, cyclic differential auxiliaries.
So we now extend the single-facet notion of `specifier equation' from Lagrange multiplier auxiliaries to this more general context.

\subsection{Discussion}

\n{\bf Remark 1} Article III's uses of {\it ((nontrivial) ab initio) consistent} carry over.  

\m 

\n{\bf Remark 2} First- and second-classness carry over within the physical Poisson sector, as does closure under Poisson brackets as a completeness criterion.  
We again postpone the possibility of second-classness to Sec \ref{Dir-Bra}, 
the TRi-Dirac Little Algorithm itself operating under the aegis that all constrains involved are first-class.  

\m 

\n{\bf Remark 3} If type 2) occurs, these first-class constraints are fed into the subsequent iteration of the algorithm. 

\m 

\n This is by, firstly, defining $\bscQ$ as one's initial $\bscP$ 
alongside the $\bscR$ subset of the candidate theory's formulation's $\bscS$ that have been discovered so far, indexed by $\fQ = \fP \coprod \fR$.

\m 

\n Secondly, by restarting from a more general form for our problem (\ref{de-for-u}) 
\beq
0     \approx     \ordial {\scQ}        
        \es       \mbox{\bf \{} \bscQ \mbox{\bf ,} \,  \ordial \cA_{\sbiu\bscQ} \mbox{\bf \}}  
        \es       \mbox{\bf \{} \bscQ \mbox{\bf ,} \,  \ordial \cH              \mbox{\bf \}}  + 
		          \mbox{\bf \{} \bscQ \mbox{\bf ,} \,  \uc{\bscQ}                   \mbox{\bf \}}     \cdot \ordial \uc{\biu}     \m . 
\label{de-for-u-2}
\eeq
Proceed recursively until one of the following termination conditions is attained. 

\m 

\n{\bf Termination Condition 0)  Immediate inconsistency} due to at least one inconsistent equation arising.

\m 

\n{\bf Termination Condition 1)  Combinatorially critical cascade}. 
This is due to the iterations of the TRi-Dirac Algorithm producing a cascade of new objects 
down to the `point on the surface of the bottom pool' that leaves the candidate with no degrees of freedom.   
This has the status of a combinatorial triviality condition.    

\m 

\n{\bf Termination Condition 2) Sufficient cascade}.  This runs `past the surface of the bottom pool' of no degrees of freedom 
into the `depths of inconsistency underneath'.

\m 

\n{\bf Termination Condition 3) Completion} is that the latest iteration of the TRi-Dirac Algorithm has produced no new nontrivial consistent equations, 
indicating that all of these have been found. 

\m 

\n{\bf Remark 4} Our input candidate set of 
generators is either itself {\it complete} 
                                                     or {\it incomplete} -- `nontrivially TRi-Dirac' -- 
													 depending on whether it does not or does imply any further nontrivial objects.
If it is incomplete, it may happen that the TRi-Dirac Algorithm provides a completion, by only an {\it combinatorially insufficient cascade} arising, 
from the point of view of killing off the candidate theory.  
														 
\m 														 

\n{\bf Remark 5} So, on the left point of the trident, Termination Condition 3) gives a Closure acceptance condition 
for an initial candidate set of constraints alongside the cascade of further objects emanating from it by the TRi-Dirac Algorithm.   	
I.e.\ one demonstrates a `TRi-Dirac completion' of the incipient candidate set of constraints.  											 

\m  

\n{\bf Remark 6} On the right point of the trident, Termination Conditions 0) and 2) are rejections thereof. 

\m 

\n{\bf Remark 7} On the final middle point of the trident, Termination Condition 1) remains the critical edge case.

\m 

\n{\bf Remark 8} In detailed considerations, clarity is often improved by labelling each iteration's $\bscR$ and $\bscQ$ by the number of that iteration.
In the case of completion being attained, (the final $\bscR$) = $\bscS$ itself -- all the first-class secondary constraints -- 
whereas $\bscQ = \bscF$: all the first-class constraints.   

\m 

\n{\bf Remark 7} We now have enough space to comment that, firstly, Dirac's own description of specifier equation was \cite{Dirac} `imposes a condition'. 

\m 

\n Secondly, that the term `fixing equations', as in e.g.\ `lapse fixing equation', is often used for in Numerical Relativity. 
C.f.\ maximal and constant mean curvature lapse fixing equations in this context.  
This useage is however a subcase of gauge-fixing, nor does all gauge-fixing involves specification of Lagrange multipliers.
E.g.\ Lorenz gauge need not be interpreted in this way. 
On these grounds, the distinct name `specifier equations' is used in this Series.

\subsection{Each iteration's problem is a linear system}

{\bf Remark 1} (\ref{de-for-u}) or its subsequent-iteration generalization (\ref{de-for-u-2}) is, once again, a linear problem.  

\m 

\n Its general solution -- now a cyclic differential -- thus splits according to 
\be 
\ordial \biu  =  \ordial \bip + \ordial \biC                                                 \m , 
\ee
for particular solution $\ordial \bip$ and complementary function $\ordial \biC$.

\m 

\n By definition, $\ordial \biC$ solves the corresponding homogeneous equation  
\beq
\ordial \uc{\biC} \cdot \mbox{\bf \{} \uc{\bscC}  \mbox{\bf ,} \,  \bscP \mbox{\bf \}} \approx 0               \m . 
\eeq
$\biC$ furthermore has the structure 
\be 
\ordial \uc{\biC} =\ordial \uc{\bic} \, \uc{\uc{\biR}}                                                       \m . 
\ee 
The $\ordial \bic$ here are the totally arbitrary coefficients of the independent solutions, 
whereas $\biR$ is again a mixed-index and thus in general rectangular matrix, just as it was in Article III.
Our general solution's destintion is to be substituted into the total differential almost-Hamiltonian, updating it.

\subsection{TRi-Dirac appending of cyclic differentials. ii)}

\n{\bf Definition 1} {\it particular-primary differential-almost-Hamiltonian} (alias {\it Dirac-type primed differential-almost})

\n{\it -Hamiltonian} is 
\be 
\ordial \cA_{\sbip\btcP}  :=  \ordial \cA^{\prime}  
                          :=  \ordial \cH            +  \ordial \uc{\bip} \cdot \uc{\bscP}                                   \m .
\ee  
\n{\bf Definition 2} The {\it differential-almost-Hamiltonian with first-class constraints appended} 
(alias {\it Dirac-type extended differential-almost-Hamiltonian}) is 
\be 
\ordial \cA_{\btcF}  :=  \ordial \cA_{\sE}  
                    :=  \ordial \cH  +  \ordial \uc{\biu} \cdot \uc{\bscP}  +  \ordial \uc{\bia} \cdot \uc{\bscS}            \m .
\ee
These $\d \bia$ are arbitrary functions, and these $\bscS$ are specifically first-class secondary constraints.
The description leaves it implicit that those auxiliaries which can be solved for, are solved for.    

\m  

\n{\bf Remark 1} Such a notion could clearly be declared for each iteration of the TRi-Dirac Algorithm, 
with the above one coinciding with the TRi-Dirac Little Algorithm attaining completeness. 
In this sense, $\ordial \cA_{\btcF}$ is itself a candidate theory's {\it maximally} extended differential-almost-Hamiltonian.  
This places order-theoretic content in the TRi-Dirac multiplicity of differential almost-Hamiltonians. 
It is $\ordial \cA_{\btcF}$ that supplants the GR extended Hamiltonian, itself a truer name for the notion that most of the literature calls `GR total Hamiltonian'.

\subsection{Removing second-class constraints}\label{Dir-Bra}

Suppose second-class constraints arising at some iteration in the TRi-Dirac Algorithm. 
Three different approaches to this are as follows.\footnote{Given Article III's motivation of forming a purely first-class constraint system, 
one might accompany some such procedures by gauge-fixing specifiers, or extending to remove the presence of specifiers.}

\m

\n{\bf Procedure A)} Remove these by replacing the incipient Poisson brackets with Dirac brackets (\cite{Dirac} and already covered in Article III).  

\m 

\n{\bf Procedure B)} Extend $\Phase$ with further auxiliary variables so as to `gauge-unfix' second-class constraints into first-class ones \cite{BT91, HTBook}. 

\m 

\n{\bf Procedure C)} Remove the objects in question by Lagrangian-level reduction.

\m 

\n{\bf Universality criterion 1} Whereas procedures A) and B) are both in principle systematically available, 
C) is not, though it is solvable for this Series of Articles's RPM and SIC examples.    
Second-class constraints can moreover always in principle\footnote{This statement follows \cite{HTBook}, though we have added the caveat `locally'  
out of gauge-fixing conditions not in general themselves holding globally.\l{foot}} 
be handled locally by thinking about them instead as `already-applied' gauge fixing conditions that can be recast as first-class constraints by adding suitable auxiliary variables.   
By this procedure, a system with first- and second-class constraints {\it extends} to a more redundant description of a system with just first-class constraints. 

\m 

\n{\bf Remark 1} As regards A), suppose second-class constraints are present at some iteration in the TRi-Dirac Algorithm.

\m 

\n{\bf Structure 2} The preceding can moreover happen on subsequent iterations of the TRi-Dirac Algorithm, were these to reveal more second-class constraints.  
I.e.\ while still in the process of investigating a physical theory's constraints, one does not yet know which are first-class. 
This is because a given constraint may close with all the constraints found so far but {\sl not} close with some constraint still awaiting discovery. 
Thus one's characterization of constraints needs to be updated step by step until either of the following apply.

\m 

\n The notion of {\it final classical bracket} alias {\it maximal Dirac bracket} thus also carries over as already-TRi. 
%

\subsection{Various notions of gauge}

\n{\bf Remark 1} Some constraints are regarded as gauge constraints. 
In general, however, exactly which kinds of constraints these comprise remains disputed in the literature.

\m 

\n{\bf Remark 2} One point agreed upon is that a Gauge Theory has an associated group $\lFrg$ of transformations that are held to be unphysical.  
The above-mentioned disjoint auxiliary variables often constitute the generators of such a group.      

\m 

\n{\bf Remark 3} Another point that is agreed upon is that second-class constraints are not gauge constraints; 
all gauge constraints use up two degrees of freedom.  

\m 

\n{\bf Dirac's Conjecture} \cite{Dirac} is that, a fortiori, all first-class constraints are gauge constraints. 

\m 

\n By this, using up two degrees of freedom would conversely imply being a gauge constraint.

\m 

\n Sec \ref{CC-Examples} however details that Dirac's conjecture has long been known to be false \cite{HTBook}.

\m 

\n{\bf Remark 4} What the gauge group acts upon is another source of diversity.

\m 

\n{\bf Definition 1} `Gauge Theory' in Dirac's sense \cite{DiracObs, Dirac} applies to {\sl data at a given time}. 
A true-name for gauge in this case is thus {\it data-gauge}. 

\m 

\n{\bf Definition 2} `Gauge Theory' in Bergmann's sense \cite{Bergmann61} applies to {\it data along whole paths}, i.e.\ trajectories in spacetime.  
A true-name for gauge in this case is thus {\it path-gauge}. 

\m 

\n{\bf Remark 5} One may extend the first of these to a fork between timeless configurations, 
configuration--velocity, configuration-change, and phase space versions of data. 

\m 

\n{\bf Remark 6} One may extend the second to make further distinction between paths and histories; 
see Part II of \cite{ABook} for details on all of these distinctions.  

\m

\n{\bf Remark 7} Make careful distinction between different {\sl notions of} Gauge Theory as here,  
and the more familiar issue of making particular {\sl choices of} gauge within the one notion of gauge, such as working in Lorenz gauge for Electromagnetism.  

\m 

\n{\bf Remark 8} In the current Article, `gauge' is meant in Dirac's    `data-gauge' sense, as is appropriate to canonical approaches 
                            see Article X for its use     in Bergmann's `path gauge' sense in spacetime approaches.  

\m 

\n{\bf Remark 9} Sec \ref{CC-Examples} and Article X moreover contain further GR or gravitational theory specific issues with the extent to which Gauge Theory ideas 
permeate into, or suffice for, Gravitation.

\subsection{Discussion}

\n{\bf Remark 1} Each of procedures A) to C) render it clear that whether a theory exhibits second-class constraints is in fact a formulation-dependent statement.  
As such, the current Series' previous mentions of `{\sl formulations of} theories' in connection to sets of constraints are indeed not superfluous.

\m 

\n{\bf Remark 2} {\it Gauge-fixing conditions} 
\be 
\bcalX \m \mbox{ with components } \m {\cal X}_{\sfX}
\ee 
may be applied to whatever Gauge Theory (for all that final answers to physical questions are required to be gauge-invariant). 

\m 
 
\n{\bf Remark 3} Of relevance to footnote \r{foot}, the square in Fig \ref{LH-Square} does not in general commute (see e.g.\ \cite{IY05}). 
%
{            \begin{figure}[!ht]
\centering
\includegraphics[width=0.6\textwidth]{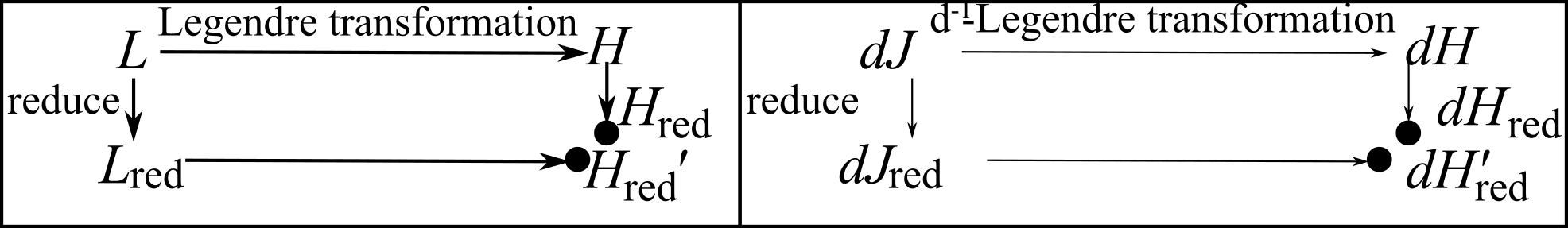}
\caption[Text der im Bilderverzeichnis auftaucht]{         \footnotesize{The in general noncommuting square of 
Legendre transformations (horizontal) and reductions (vertical) in a) plain and b) TRi settings. }}
\label{LH-Square}\end{figure}            }

\m 

\n{\bf Remark 4} Also as regards footnote \r{foot}, at least in the more standard theories of Physics, 
first-class secondary constraints can be taken to arise from variation with respect to mathematically disjoint auxiliary variables.
The effect of this variation is to additionally use up part of an accompanying mathematically-coherent block of variables 
that otherwise contains partially physical information.

\subsection{TRi Dirac Full Algorithm}

Proceed as in Sec \ref{TRi-DA}, except that whenever second-class constraints appear, one switches to (new) Dirac brackets that factor these in.  
This amounts to the possibility of a fifth type of equation, in parallel with Sec III.2.11.    

\m 

\n{\bf Type 4)} {\bf Further second-classness} may arise.

\m 

\n{\bf Aside 1} Let us distinguish between auxiliaries used for appending and smearing variables.
The latter are more widely applicable since their job -- `multiplication by a test function' -- 
is to render rigorous a wider range of `distributional' manipulations provided that these occur within an integral.    
In particular, this applies to classical Field Theories' Assignment of Observables (for which there is no appending procedure). 

\m 

\n{\bf Aside 2} For later convenience, we express the $\bFlin$ in manifestly homogeneous linear form:
\beq
\Flin_{\sfN}  =  {\cal F}[\bfQ]^{\sfA}\mbox{}_{\sfN}\fP_{\sfA}  \m .
\label{Flin-Form}
\eeq
This includes the possibility of ${\cal F}$ being differential operator-valued so as to accommodate Electromagnetism, Yang--Mills Theory and GR. 
				
\m 

\n{\bf Remark 1} At each iteration, then, one ends up with a bare differential Hamiltonian with first-class constraints appended using cyclic differentials.  
The final such is once again denoted by $\d \cA_{\btcF}$, corresponding to having factored in all second-class constraints and appended all first-class constraints.  
Each other notion of differential almost Hamiltonian above can also be redefined for Dirac brackets, whether maximal or at any intermediary stage.   

\m 

\n{\bf Remark 2} As the endpoint of our elaboration of `extended Hamiltonians' along Dirac's lines, on the one hand, and TRi on the other, 
we accord this object it a more compact final name for future reference, namely `Rid-amiltonian'.  
On the one hand, this builds in that this is not just TRi but Ri: Relationalism implementing. 
On the other hand, `amiltonian' is short for an almost-Hamiltonian 
(c.f.\ transforming between Hamiltonians sometimes being phrased as from a Hamiltonian to a transformed Kamiltonian). 
`d-amiltonian' is, similarly, short for a differential-almost-Hamiltonian. 
We preserve our notation `d A' notation for this final concept, by using   
\be 
\ordial \cA_{\sR\si}  \m   
\ee 
and that e.g.\ ${\cal S}_{\sJ}$: is Jacobi action, so the subscript comes first in the corresponding naming. 
This is technically a three-aspect object: a Ri-object that additionally belongs to the machinery of testing for Closure. 
It is not however yet a three-aspect-{\sl incorporating object}; if it were, its subscript would be `CC-Ri'; 
it is only promoted to this stage when the TRi-Dirac Algorithm has confirmed its pertinence to a consistent theory.  

\m 

\n{\bf Type 5} {\bf Discovery of topological obstructions} also carries over mathematically unaltered to the TRi setting.
We present a more extensive discussion of such matters in Article XIV.

\vspace{10in}
 
\section{Examples of distinctions between types of constraint}\label{CC-Examples}

Let us next justify the finer distinctions between types of constraint made in Article III.     

\m 

\n{\bf Example 1)} The constraints considered so far in this Series of Articles -- in particular RPM's          $\u{\bscP}$, $\u{\sbcL}$, $\scD$, $\scE$, 
                                                                                   Electromagnetism's          $\scG$, 
																				   Yang--Mills Theory's        $\scG_I$, 
																				   and GR-as-Geometrodynamics' $\u{\bscM}$, $\scH$ -- are all first-class. 
It is thus useful to now provide examples of {\bf second-class constraints}, so that readers see that these do in fact reside in some familiar theories 
which are either standard observationally substantiated theories, or just one step therefrom.   

\m 

\n i) In the $\biQ = (\mA_i, \Phi)$ formulation of the `massive analogue of Electromagnetism' (alias {\it Proca Theory}),   
\beq
\scC  :=  \bpa \cdot \bE + m^2 \Phi  
       =  0                          \m .  
\label{C-Proca}
\eeq
This indeed uses up only one degree of freedom, so this theory has one more physical mode than Electromagnetism itself 
(from two transverse-traceless modes to having a longitudinal mode as well).   

\m 

\n ii) Specifically Gravitational Theories with second-class constraints include {\sl Einstein--Dirac Theory} (i.e.\ GR with spin-1/2 fermion matter) \cite{DEath} 
                                                                             and {\sl Supergravity}                                          \cite{SqrtTeitelboim}.  

\m 

\n For the first four theories above, the absence of second-class constraints means that the Dirac chain consists of just the incipent Poisson bracket itself. 
I.e. the single-element chain, for which the bottom and top elements coincide, so the maximal Dirac bracket is just this case's incipient Poisson bracket. 

\m 

\n For Proca Theory, however, precisely 1 step in the Dirac(-type) Algorithm produces a second-class constraint, 
so the Dirac chain consists of the incipient Poisson bracket followed by the final maximal Dirac bracket.   

\m 

\n We finally leave finding the simplest and most mundane examples of Dirac chains with nontrivial middle as an exercise for the readers. 

\m 

\n{\bf Example 2)} {\bf Relational recovery of Gauge Theory} (in Dirac's data sense). 
With Configurational Relationalism's $\lFrg$ being a candidate group of physically irrelevant motions, 
in general it remains to be ascertained whether the $\bShuffle$ provided by Best Matching is a gauge constraint $\bGauge$ which corresponds to $\lFrg$.

\m 

\n{\bf Remark 1} Whether there is $\lFrg$ compatibility can at least in part be investigated prior to consideration of constraints.

\m 

\n This is since, on the one hand, 
\be 
\mbox{\bf\{} V \mbox{\bf,} \, P_{\Frg} \mbox{\bf\}}
\ee 
in 
\be
\mbox{\bf\{} \Chronos \mbox{\bf ,} \, P_{\Frg} \mbox{\bf \}}
\ee  
can already be examined prior to constraints: adopting a $\lFrg$ comes with Equipping with Brackets. 

\m 

\n On the other hand, one does not assess $\FrT(\FrQ, \dot{\FrQ})$ itself, 
which is tied to constraints being more simply and systematically handled in Hamiltonian-type formulations.

\m 

\n{\bf Remark 2} The ensuing action can be viewed as a map from a structure that is a fibre bundles twice over: both a tangent bundles {\sl and} $\lFrg$-fibre bundles.  
Specifically, it is 
\be 
\bFrP  \big(  \FrT(\FrQ), \lFrg  \big)  \m ,
\ee 
rather than         
\be 
\FrT  \big(  (\bFrP(\FrQ, \lFrg)  \big)
\ee 
due to the nontrivial part of the $\lFrg$ action being on the tangent bundles' fibres.
This being a $\lFrg$-fibre bundle mathematically can moreover require excision of certain degenerate configurations \cite{A-Killing,A-Cpct},     
which in turn is not a relationally bona fide procedure.   

\m 

\n{\bf Counter-example 3)} Despite {\bf Dirac's Conjecture}, 
\be 
\bFlin \centernot{\Rightarrow} \bGauge
\ee 
by e.g.\ the following technically constructed but not physically motivated counter-example given by Henneaux and Teitelboim \cite{HTBook}.  
The Lagrangian 
\be 
L = \half \, \mbox{exp}(y)\dot{x}^2
\ee
gives a constraint 
\be 
p_x = 0
\ee 
which is first-class but not associated with any gauge symmetry.

\m 

\n{\bf Example 4)} Whereas $\u{\sbcL}$, $\sbcG$, $\u{\bscM}$ are uncontroversially gauge constraints, 
the gauge status of $\scH$ and even $\scE$ remain disputed.    
I.e.\ $\Chronos$ {\bf constraints entail `gauge subtleties'}.
Some arguments of note in this regard have been given by \K, Barbour and Foster \cite{K93, K99, BF08}. 

\m 

\n This point is, moreover, directly at odds with \cite{HTBook}, which {\sl transform} to and from constraints of the form $\Quad$. 
The Author pointed out \cite{AObs} that this discrepancy is due to the following. 

\m 

\n On the one hand, \cite{HTBook} allowing for $t$-dependent canonical transformations, $Can_t$ (see Appendix \ref{Mor}). 

\m 

\n On the other hand, the relational whole-universe context has no primary-level $t$, by which it is not licit to adopt $Can_t$ in this worldview.  
$\Chronos$ and $\bGauge$ are consequently qualitatively distinct in the relational context.  
The relational context furthermore makes distinction between Constraint Providers for, firstly, $\bShuffle$ candidates for $\bGauge$, 
                                                                                 and, secondly, $\Chronos$.  

\m 

\n{\bf Remark 3} It is fitting for {\sl Configurational} Relationalism to be associated with a {\sl data}-gauge notion. 

\m 
																				 
\n{\bf Remark 4} GR's $\scH$ is moreover a case study into the extent to which Gauge Theory ideas permeate into, or suffice for, Gravitation.  

\m 

\n i) To what extent is one still dealing with Gauge Theory when groupoid rather than group structure is present?
For inclusion of $\scH$ means that one's first-class constraints locally form not an algebra but an algebroid: the Dirac Algebroid.  

\m 

\n ii) To what extent can `hidden symmetries' can be treated as gauge symmetries?  
This is relevant to $\scH$ since this encodes Refoliation Invariance as a hidden symmetry.  

\m 

\n iii) Pons, Salisbury and Sundermeyer \cite{PSS09, PSS10} argue for GR's evolution generator mimicking, rather than being, a gauge generator; 
see Article XII for more about this.

\m 

\n Constraint algebroids and hidden symmetries moreover enter Gravitational Theory beyond GR; see e.g.\ Sec \ref{CC-Ex}. 
																				 
\m 

\n{\bf Counter-example 6)} Article XI moreover presents a SIC example of 
\be 
\bGauge \centernot{\Rightarrow} \bFlin \m , 
\ee
i.e.\ failure of a partial converse to Dirac's Conjecture.    

\m 

\n This discussion, by delving into one or both of Gravitational and Background Independent ventures,
takes us in a direction considerably outside of the scope of Henneaux and Teitelboim's excellent book \cite{HTBook}.

\section{Constraint algebraic structures}\label{NoC}

\subsection{Overview}

{\bf Structure 1} The end product of a successful candidate theory's passage through the Dirac Algorithm 
is a {\it constraint algebraic structure} consisting solely of first-class constraints closing under Poisson (or more generally Dirac) brackets. 

\m 

\n{\bf Remark 1} Article III already covered the general form of this.   

\m 

\n{\bf Structure 2} We now furthermore identify individual constraint algebraic structures to each be a {\it Poisson algebraic structure} in the obvious sense; 
see e.g. \cite{Gengoux} for an introduction to Poisson algebras.  

\m 

\n{\bf Structure 3} The end product of a successful candidate theory's passage through the Dirac Algorithm is a {\it constraint algebraic structure} 
consisting solely of first-class constraints closing under Poisson (or more generally Dirac) brackets. 
This is already-TRi and so carries over.  

\m 

\n{\bf Structure 4} Sec III.2.17's lattice of subalgebraic structures is an already-TRi structure as well, and so also carries over.

\subsection{Examples of constraint algebraic structures}\label{Con-Alg-Str}

Cases with reduction at any classical level explicitly attained are aided in the matter of closure by one or both of the following means.

\m 

\n{\bf Means a)} Having fewer constraints to form brackets out of.

\m 

\n{\bf Means b)} Making use of single finite-theory classical constraints always Abelianly closing with themselves by symmetric entries into an antisymmetric bracket.  

\m 

\n{\bf Example 0)} Minisuperspace has just one constraint, $\Chronos = \scH_{\sm\si\sn\si}$, so we have 
\be
\mbox{\bf \{} \scH_{\sm\si\sn\si} \mbox{\bf ,} \, \scH_{\sm\si\sn\si} \mbox{\bf \}}  =  0   \m : 
\ee
closure as a single-generator Abelian Lie algebra. 
It thereby passes all kinds of Constraint Closure.

\m 

\n{\bf Example 1.R)} Reduced Euclidean RPM also has just one constraint, $\Chronos = \w{\scE}$, so we once again attain Constraint Closure in the algebraic form 
\be 
\mbox{\bf \{} \w{\scE} \mbox{\bf ,} \, \w{\scE} \mbox{\bf \}}  =  0  \m .  
\label{we,we}
\ee 
This has the added merit that Configurational Relationalism has been incorporated (minisuperspace did not have any of this to begin with).  

\m 

\n{\bf Example 1.U)} Unreduced Euclidean RPM has the constraint algebra described in Sec III.2.18.  

\m 

\n Reduced similarity RPM obeys (\ref{we,we}) for redefined $\w{\scE}$ as well.  
Note that explicit reduction is easy in 1-$d$, solved with a bit of geometry in 2-$d$, 
and yet manifests much harder topology and geometry for $\geq 3$-$d$, with solutions at best local in $\w{\FrQ}$.  

\m

\n These RPM models can be summarized by 
\beq
\mbox{RPMs realize the} \m \m  \{ \scE \}    \, \times \, \bFrG\mbox{\bf auge} \m \mbox{ subcase of } 
                                 \m   \bFrC\mbox{\bf hronos} \, \times \, \bFrG\mbox{\bf auge}                     \m .  
\label{RPM-Direct}    
\eeq
\n The final Rid-amiltonian for Euclidean RPM is  
\be
\d \mA_{\btcF} \:=  \d I  \,  \scE   +      \d \u{B} \cdot \u{\scL}     \mbox{ (strictly $\m + \d\u{\lambda} \cdot \u{P}^B$) }       \m . 
\ee 
\n{\bf Example 2)} Electromagnetism has Gauss constraint $\scG(\u{x})$; 
not by itself admitting a TRi formulation anyway, we smear this in the usual manner with scalar functions $\zeta(\u{x}), \omega(\u{x})$. 
The resulting constraint algebra is 
\be 
\mbox{\bf \{} ( \, {\scG} \, | \, \zeta \, )  \mbox{\bf ,} \, ( \, {\scG} \, | \, \omega \, ) \mbox{\bf \}}  =  0  \m . 
\ee 
This of course reflects that the underlying gauge group is the Abelian $U(1)$.  
$( \m | \m )$ is here the integral-over-flat-space functional inner product.  
The final Hamiltonian for Electromagnetism is 
\be 
\cH_{\tcF}  =  \cH  +  \Lambda \scG  \mbox{ (strictly $\m + \lambda \cdot \Phi$) }       \m . 
\ee
\n{\bf Example 3)} Yang--Mills Theory has Gauss constraint $\scG_I(\u{x})$. 
Once again not by itself admitting a TRi formulation anyway, 
we smear this in the usual manner albeit now with internal-vector functions $\zeta^I(\u{x}), \omega^I(\u{x})$.
The resulting constraint algebra is 
\be 
\mbox{\bf \{} ( \, {\scG}_I \, | \, \zeta^I \, )  \mbox{\bf ,} \, ( \, {\scG}_J \, | \, \omega^J \, ) \mbox{\bf \}}  \es  {G^K}_{IJ} ( \, \scG_K \, | \, [\zeta, \omega]^K )  \m . 
\ee 
for structure constants $\u{\u{\u{\biG}}}$ and internal-index commutator Lie bracket $[ \m , \m ]$.   
This of course represents the corresponding Yang--Mills gauge group.
The final Hamiltonian for Yang--Mills Theory is
\be 
\cH_{\tcF}  =  \cH  +  \Lambda^I \cdot \scG_I  \mbox{ (strictly $\m + \lambda^I \cdot \Phi_I$) }       \m .
\ee

\subsection{Example 4) Full GR}\label{GR-Ex}

\n{\bf Structure 1} Since GR does admit a TRi reformulation, we perpetuate this by adopting a TRi smearing: 
vectors $\pa \u{L}$ and $\pa \u{M}$ and scalars $\pa J, \pa K$.  

\m 

\n{\bf Structure 2} In terms of this, the `TRi-dressed' version \cite{AM13, ABook} 
of the Dirac algebroid (III.53-55) \cite{Dirac51, Dirac58, Tei73} 
formed by GR's constraints (II.20--22) is 
\beq
\mbox{\bf \{} ( \u{\bscM} \, | \, \pa \u{\mL} ) \mbox{\bf ,} \, ( \u{\bscM} \, | \, \pa \u{\mM} ) \mbox{\bf \}}  \es  
              ( \u{\bscM} \, | \, \u{\mbox{\bf |[} \pa \mL \mbox{\bf ,} \pa \mM \mbox{\bf ]|}} )  \m .
\label{TRi-Mom,Mom}
\eeq
\be
\mbox{\bf \{} (    \scH    \, | \,    \pa \mK     ) \mbox{\bf ,} \,(    \u{\scM}  \, | \,    \pa \u{\mL}    ) \mbox{\bf \}}  \es  
              (    \pounds_{\pa \underline{\sL}} \scH    \, | \,    \pa \mK    ) \m , 
\label{TRi-Ham,Mom}
\ee
\be 
\mbox{\bf \{} (    \scH    \, | \,    \pa \mJ    ) \mbox{\bf ,} \, (    \scH    \, | \,    \pa \mK    )\mbox{\bf \}}  \es    
(   \u{\scM} \cdot \u{\u{\mh}}^{-1} \cdot  \, | \,    \pa \mJ \, \overleftrightarrow{\u{\pa}} \pa \mK    ) \m . 
\label{TRi-Ham,Ham}
\ee
$( \m | \m )$ is now the integral-over-curved-space functional inner product, and 
$[ \m , \m ]$ the differential-geometric commutator Lie bracket.  

\m 

\n For GR, the final Rid-amiltonian is           
\be 
\pa \cA_{\tcF} = \pa \mI   \, \cH   +   \pa \u{\mF} \cdot  \u{\scM} \mbox{ (strictly $\m + \pa \u{\uplambda} \cdot \u{\mp}^{\upbeta}$)}  \m .       
\ee 
\n{\bf Remark 1} This resmearing does not however change any of Article III's interpretational comments;  
in particular, GR-as-Geometrodynamics succeeds in attaining Constraint Closure at the classical level.
{\sl The above TRi-dressed form for this, moreover, culminates classical GR-as-Geometrodynamics' 
      consistent incorporation of the first three Background Independence aspects, 
alias consistent overcoming    of the first three Problem of Time         facets.}  
Let us celebrate by reissuing Fig III.1.f-h) in final TRi-smeared form: Fig \ref{Teit-Pic-7}.
%
{            \begin{figure}[!ht]
\centering
\includegraphics[width=1.0\textwidth]{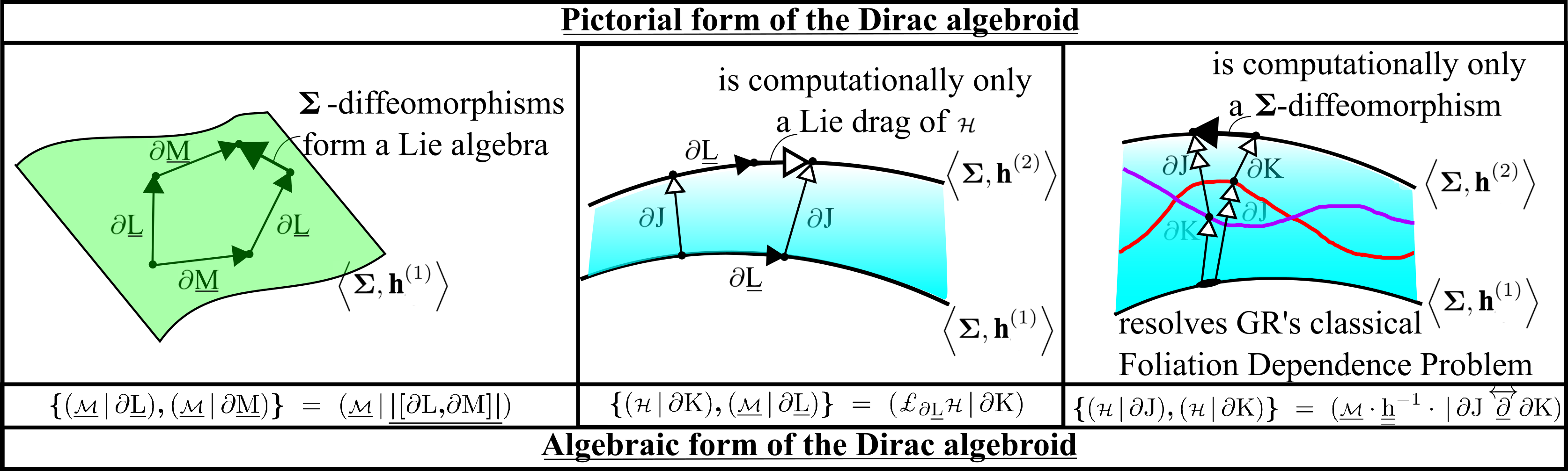} 
\caption[Text der im Bilderverzeichnis auftaucht]{ \footnotesize{ 
TRi-dressed Dirac algebroid of GR constraints' a) geometrical significance and b) algebraic structure.  } }
\label{Teit-Pic-7} \end{figure}          }

\m 
 
\n{\bf Remark 2} Some further remarks about the Dirac algebroid not yet made in this Series are as follows. 
It already features in Minkowski spacetime $\mathbb{M}^n$ in {\sl general coordinates} so as to model fleets of accelerated observers therein. 
This is in fact the context in which Dirac first found this algebroid \cite{Dirac51} though he subsequently considered the GR case in \cite{Dirac58}. 

\m 

\n{\bf Remark 3} That the Poisson bracket of $\Chronos = \scH$ with itself (\ref{TRi-Ham,Ham}) gives rise to $\bShuffle$ constraints $\u{\scM}$.
This indicates a greater amount of `togetherness' between Temporal and Configurational Relationalism than the RPM model arena exhibits.  
Consequently, in the GR setting, Temporal Relationalism cannot be entertained without Configurational Relationalism. 
This is in contrast with how the two can be treated piecemeal in RPM.  
As an integrability, this is analogous to Thomas precession.

\m 

\n{\bf Structure 3} $SO(3, 1)$ decomposes under rotations--boosts -- $\u{J}$--$\u{K}$ split -- schematically as  
\beq
\mbox{\bf |[}\u{J} \mbox{\bf ,}  \, \u{J}\mbox{\bf ]|} \sim \u{J}     \mbox{ } , \mbox{ } \mbox{ } 
\mbox{\bf |[}\u{J} \mbox{\bf ,}  \, \u{K}\mbox{\bf ]|} \sim \u{K}     \mbox{ } , \mbox{ } \mbox{ } 
\mbox{\bf |[}\u{K} \mbox{\bf ,}  \, \u{K}\mbox{\bf ]|} \sim \u{K} + \u{J}        \mbox{ } . 
\eeq
The last bracket is key, since by this the boosts $\u{K}$ do not constitute a subalgebra.\index{boosts}
Thomas precession then refers to the rotation arising in this manner from a combination of boosts.

\m 

\n This is of course a spacetime generator matter rather than a constrained one; 
it is included here, rather, for its analogy with GR's constraints' Dirac algebroid. 

\m 

\n This is moreover a case in which linearly recombining the two blocks reveals a simpler split form, in accord with the 
\be 
so(3, 1) \cong so(3) \times so(3)
\ee 
accidental relation, as is well-known in both Group Theory and Particle Physics. 

\m 

\n This does however amount to abandoning one's originally declared partition of generators.
This partition is of no importance in the current example, but the corresponding partition in the GR case is often considered to be significant.  

\m 

\n{\bf Structure 4} The GR constraints analogy with Thomas Precession is as follows.  
\beq
\mbox{GR manifests the } \m \{\scH\}     \, \Thomas \, \{ \u{\scM} \} \m \mbox{ subcase of } 
                         \m \bFrC\mbox{\bf hronos} \, \Thomas \, \bFrG\mbox{\bf auge}  \m .  
\label{GR-Thomas}    
\eeq
So, there is a parallel between composing two boosts         producing a rotation: Thomas precession,   
                           and composing two time evolutions producing a spatial diffeomorphism: {\it Moncrief--Teitelboim on-slice Lie dragging} \cite{MT72}.

\m 						   
						   
\n{\bf Limitation 1} on this analogy is that the GR version takes the form of an algebroid, as required to encode the multiplicity of foliations. 

\m 

\n{\bf Limitation 2} is that, unlike for Thomas Precession, the integrability cannot be undone by linearly combining constraints. 

\m 

\n [There is however a matter time approach supporting redefined constraints that close as a Lie algebra; 
see Part II of \cite{ABook} for downsides to matter time approaches however.]

\m 

\n{\bf Remark 4} Aside from minisuperspace's collapse to an Abelian Lie algebra (III.56), other simpler subcases of note are as follows. 

\m 
 
\n{\bf Example 5} Strong Gravity \cite{I76} demonstrates \cite{San} a smaller a collapse in which both the integrability and the algebroid nature are lost; 
this is covered in Article IX

\m 

\n{\bf Example 6} In spatial dimension 1 and $\mathbb{S}^1$ topological manifold, the Dirac algebroid collapses to a Lie algebra (albeit infinite-$d$)
that is well known: the Witt algebra, or, with central extension, the Virasoro algebra \cite{GKO}.  

\m 

\n This simplification does not however extend to GR in spatial dimension 2 or higher. 

\m 

\n{\bf Remark 5} Upon including minimally-coupled matter (including no curvature couplings), 
one has the {\it Teitelboim split} \cite{Teitelboim} for minimally-coupled matter 
\beq
\scH = \scH^{\sg} + \scH^{\Psi} \mma 
\eeq
\beq 
\u{\scM} = \u{\bscM}^{\sg} + \u{\bscM}^{\Psi}   \m . 
\label{Teitelboim-Split}
\eeq 
\n Teitelboim \cite{Teitelboim} moreover showed that gravitational and minimally-coupled matter parts obey the Dirac algebroid {\sl separately}.
This follows from (VI.90), the general form taken by minimally-coupled matter potentials, and (\ref{Teitelboim-Split}).  

\m 

\n There is no difficulty with extending this approach to Einstein--Maxwell or Einstein--Yang--Mills theories; 
see \cite{Teitelboim} for a non-TRi account; 
for a TRi version, just feed Article VI's constraints with TRi-smearing into the TRi-Dirac Algorithm.  

\m 

\n This immediately extends to scalar Gauge Theories as well. 
For fermionic gauge theories, one needs to work with beins (or similar), by which frame constraints enter at the secondary level.  
This does not however change the integrability structure or algebroid nature of the subsequent algebraic structure.

\m  

\n{\bf Remark 6} See \cite{BojoBook} for a further brief introduction to the Dirac algebroid, 
and e.g.\ Appendix V of \cite{ABook} for a brief introduction to algebroids more generally.
\cite{Algebroid1, Algebroid2, LandsmanBook, GS08} are texts containing more detailed accounts on the latter.
The Author strongly suspects both the latter and the former to be in the infancy of their developments as academic disciplines.

\section{Constraint Closure itself}\label{CC}

This is also already-TRi and so carries over from Article III. 
The same applies to Article III's discussion of Constraint Closure Problems, 
which moreover include facet interferences with Temporal and/or Configurational Relationalism.

\subsection{Seven Strategies for dealing with Constraint Closure Problems}

\n If a severe form of the Constraint Closure Problem strikes, one may have to entirely abandon the candidate theory's triple 
$\langle \FrT(\FrQ), \lFrg, {\cal S} \rangle$.  
I.e.\ the Machian variables, a group acting thereupon and the Jacobi--Synge geometrical action.\footnote{One might augment this 
to a quadruple by considering varying the type of group action of $\nFrg$ on $\sFrT(\sFrQ)$.}
%
In some cases, however, modifying one or more of these may suffice to attain consistency. 
This gives the 2-sided cube of strategies of Fig \ref{CC-Strat}.  
%
{            \begin{figure}[!ht]
\centering
\includegraphics[width=0.7\textwidth]{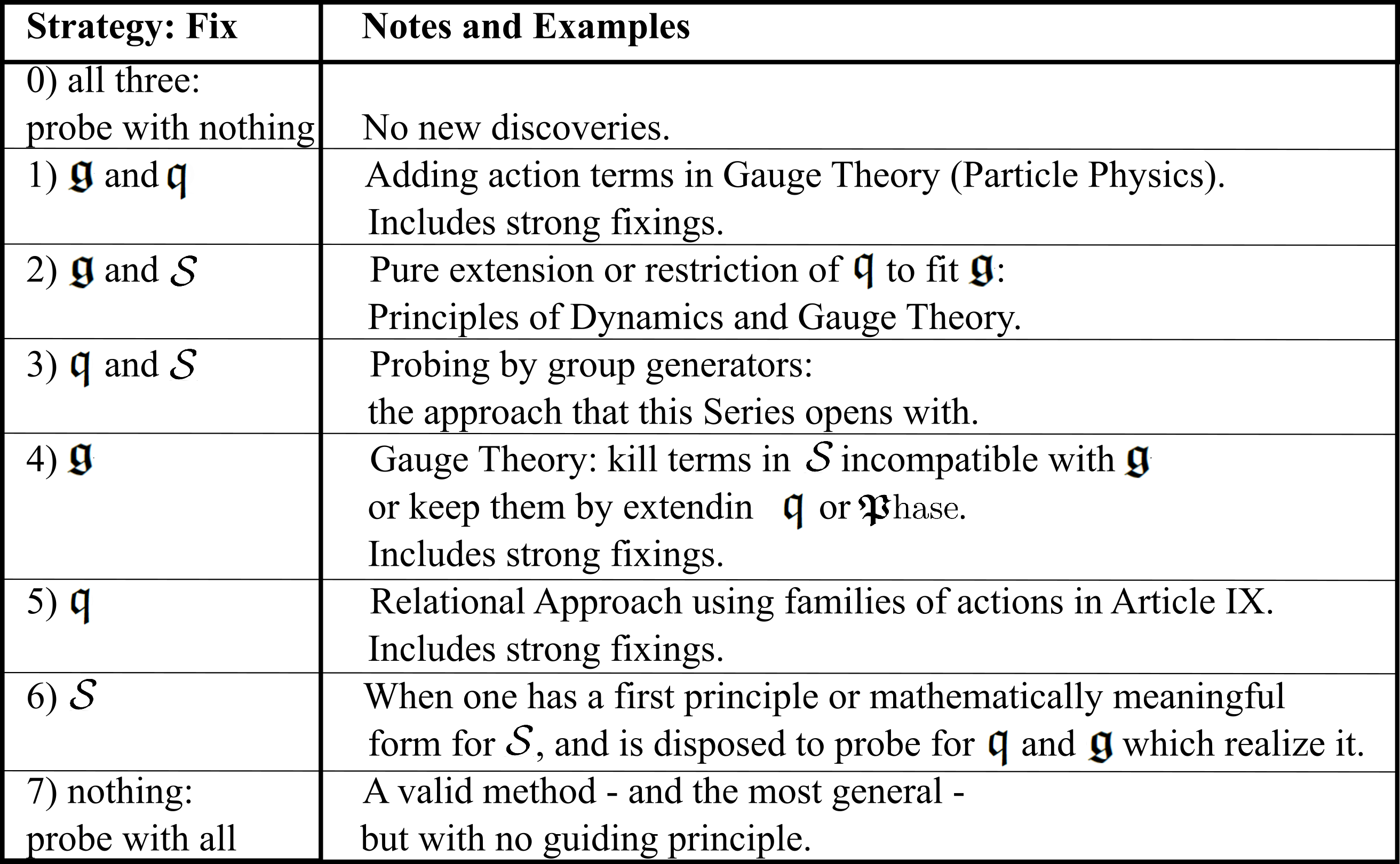} 
\caption[Text der im Bilderverzeichnis auftaucht]{ \footnotesize{Seven strategies with some capacity for 
generating new theories from what is allowed by Constraint Closure. 
In each case, the structures which remain fixed act as a guiding principle. 
[Keeping a given $\nFrg$, what physics ensues? 
What about with a given $\sFrQ$? 
A given ${\cal S}$?]
The unfixed complement structures correspond to types of probing for new theories.  
Note how this reasoning pitches the Relational Approach as a complementary method to Gauge Theory.
Also, paralleling how Gauge Theory can be attempted with extra terms in the action which are then ruled out by lack of $\nFrg$ compatibility, 
the current Series' `Relational Approach' comes in a larger version in which whole families of candidate theories are treated at once (Article IX).} }
\label{CC-Strat} \end{figure}          }

\m

\n{\bf Remark 1} Fig \ref{CC-Strat}'s strategic diversity continues to apply if $\Phase$ and an integrated 
($\ordial$A-)Hamiltonian -- or its constituent set of constraints in whole-universe theories -- are considered in place of $\FrQ$ and ${\cal S}$.  
Similar considerations apply in spacetime formulations of ${\cal S}$ with $\lFrg_{\sS}$ acting thereupon (see Article X) 
and at the quantum level (further extending the Hamiltonian presentation).  

\m 

\n{\bf Remark 2} Preserving a particular $\lFrg$ in Particle Physics includes insisting on a particular internal gauge group, 
or on the Poincar\'{e} group of SR spacetime.  

\m 

\n{\bf Remark 3} {\it Strong vanishing} involves fixing hitherto free constants in ${\cal S}$ so as to avoid the problem.

\m 

\n{\bf Remark 4} Among the figure's three entities, $\FrQ$ is the one taken to have some tangible physical content. 
As such, it has the {\sl a posteriori right to reject} \cite{RWR, Phan, Lan2} a proposed $\lFrg$ by triviality or inconsistency.  

\m 

\n This role extends to using $\Phase$ instead.  
\n One can indeed also consider our eight strategies for the triple 
\be
\langle \Phase , \lFrg, {\cal A}_{\btcF} \rangle  \m ,
\ee  
where $\lFrg$ now acts on $\Phase$ and ${\cal A}_{\btcF}$ is the integrated extended $\ordial$-almost Hamiltonian. 

\m 

\n{\bf Remark 5)} One consequence of adopting strategies permitting extension or reduction of $\FrQ$ or $\Phase$ is that formulations with second-class constraints 
are ultimately seen as half-way houses to further formulations which are free thereof.
This is largely the context in which both the effective formulation and the Dirac bracket formulation were developed, 
with $\Phase$ getting extended in the former and reduced in the latter. 

\m

\n{\bf Remark 6} With reference to Article III's classification of Closure Problems, 
on the one hand whichever of Topological Obstruction, 
                             Cascade, 
				             Involvement of Specifiers, and 
				             Algebraic Interference can be addressed by any of these strategies.

\m 

\n On the other hand, Enforced Group Extension and Enforced Group Reduction require one of strategies 3) or 5-7).  

\m 

\n{\bf Remark 7} Going full circle, we remind the reader that `Cascade' includes each of relational triviality, 
                                                                                         triviality, or 
																	                     inconsistency as worst-scenario bounding subcases.  
We called the last two of these jointly `sufficient cascade', so let us use `relationally sufficient cascade' for the three cases together.

\subsection{Further realizations of Constraint Closure Problems}\label{CC-Ex}

\n The below examples serve to populate our finer distinctions between types of Constraint Closure phenomena. 
These examples would however belong more naturally in a longer account of Comparative Background Independence \cite{ABook, A-Killing, A-Cpct, A-CBI}, 
wherein their Configurational and Temporal Relationalism would have already been laid out.  

\m 

\n{\bf Example 1}) [{\sl of needing to extend} $\lFrg$] \cite{ABook}. Correcting one's action with respect to just the combination of translations $\u{P}$ 
                                                                                   and special conformal transformations $\u{K}$ fails
because the ensuing secondary constraints $\u{\bscP}$, $\u{\bscK}$ do not form a group without both scaling $\scD$ and rotations $\u{\sbcL}$.
I.e.\ schematically, 
\be 
\mbox{\bf \{} \u{\bscP}  \mbox{\bf ,} \,  \u{\bscK} \mbox{\bf \}}  \m \sim \m  \scD + \u{\sbcL}
\ee 
This additionally serves as an example of mutual integrabilities.  

\m 

\n{\bf Example 2)} [{\sl of failure of} $\lFrg$ {\sl to be a gauge group}] This is a valid problem in the absence of $\scH$, 
so attempting to impose $U(1)$ symmetry on {\sl Proca Theory} suffices.
A constraint (\ref{C-Proca}) arises, but this is second-class so it only uses up 1 degree of freedom.

\m 

\n One way out involves considering that Proca Theory rejects quotienting by $U(1)$ (Strategy 3).

\m 

\n Another, if one insists on retaining $U(1)$, is to consider $m = 0$ to arise as a strong condition (Strategy 1).
This gives a longer route to the exclusion of mass terms from $U(1)$ symmetric 1-form actions.   

\m 

\n Proca Theory can indeed be handled with by Dirac brackets {\sl or} the effective method (Exercise!). 

\m 

\n{\bf Remark 1} Both of these methods and the preceding strong condition all offer distinct minimalistic ways of 
dealing with a mismatch in an original candidate triple $\langle \lFrg, \FrQ, {\cal S} \rangle$. 

\m 

\n{\bf Example 3)} [{\sl of abandoning ship due to structural incompatibility}] Consider Best Matching with respect to the affine transformations $Aff(d)$ 
within a Euclidean-norm kinetic arc element \cite{AMech}. 
This produces a constraint $\scE$ which is incompatible with $Aff(d)$. 
This defect can be traced from $\scE$ possessing a Euclidean norm back to the kinetic arc element assumed.
In this case, however, progress is not via extension or the Dirac bracket, 
but rather by acknowledging that one needs to build an arc element free from any residual Euclidean prejudices.
Thus one `abandons ship', in the sense of forfeiting a type of ${\cal S}$ for all that one can pass to a different type of working theory \cite{AMech}.
This example's required alteration so as to attain consistency is however unrelated to changing any of 
$\lFrg$, $\FrQ$ or ${\cal S}$.  

\m 

\n{\bf Example 4)} [{\sl Best Matching itself sunk by a Constraint Closure Problem.}]
Suppose we try to impose a $\lFrg$ including both the $GL(d, \mathbb{R})$ 
transformations and the special conformal transformations acting on flat space.
Their mutual bracket however forms an obstruction term \cite{AMech, ABook}. 
This sinking is underlied by Best Matching being just a piecemeal consideration of generators while Constraint Closure involves relations as well as generators.

\m 

\n{\bf Example 5)} [{\sl Mutual second-classness.}] 
We show in Article IX that if the {\sl conformogeometrodynamical conditions} $\mp = 0$ or $\mp/\sqrt{\mh} =$ const are regarded as constraints, 
they are second-class with respect to $\scH$. 
An extension strategy for this is outlined in e.g.\ \cite{MBook}; on the other hand, the Dirac brackets approach remains untried in this case. 

\m 

\n{\bf Example 6)} [{\sl of adjoining new secondary constraints, forcing $\lFrg$ to be extended via a new integrability arising.}]
In attempting to set up metrodynamics (no spatial diffeomorphisms presupposed) the Poisson bracket of two $\scH$'s continues to imply a momentum constraint 
(see Article IX).
So ab initio (\ref{GR-Thomas}) continues to arise \cite{MT72}, giving the claimed extension by integrability.  
This is furthermore an example of Article II's point that {\sl existence of a natural action of $\lFrg$ on $\FrQ$ 
does not guarantee that $\lFrg$ represents the totality of physically irrelevant transformations}.  
Our enlargement amounts to being forced to pass from $\lFrg = id$ to the $Diff(\bupSigma)$ that corresponds to $\u{\bscM}$.  
This furthermore illustrates that $\Chronos$ can have its own say as to what form (part of) the $\bShuffle$ is to take.

\m 

\n Article IX moreover also shows that such a $\lFrg$-Closure Problem does not however occur in attempting metrodynamical Strong Gravity; 
this remains consistent with just the one Hamiltonian constraint.   
%

\m 

\n{\bf Example 7)} {\sl Supergravity} exemplifies, firstly, $\bFlin$ {\sl not closing as a subalgebraic structure}.
This is by the bracket of two linear supersymmetric constraints giving the quadratic Hamiltonian constraint. 
Since the bracket of two Hamiltonian constraints still returns a linear momentum constraint, moreover, Supergravity also exemplifies {\sl two-way integrability}   
\be 
\bFrF\mbox{\bf lin} \TwoWay \bFrC\mbox{\bf hronos}  \m .
\ee

\section{Conclusion}

\n Suppose a given theory's Constraint Closure succeeds as per the current Article. 

\m 

\n 1) For such as Minisuperspace or Temporally-Relational but Spatially-Absolute Mechanics which do not realize any nontrivial Configurational Relationalism, 1)
\be
\mbox{the status of } \m {\cal S}^{\st\sr\si\sa\sll}_{\sT\sR\si} \m \mbox{ can be upgraded to } {\cal S}_{\sC\sC-\sT\sR\si} 
\ee 
\n ii) The $\Chronos$ arising from this action closes (by itself in these two examples).

\m 

\n iii) This $\Chronos$ can be rearranged to form a Machian emergent time of the conceptual form 
\be 
t^{\se\sm}_{\sC\sC\mbox{-}\sR\si}  \m . 
\ee 
Ri here embraces both Configurational and Temporal Relationalism.  

\m 

\n For such as Electromagnetism or Yang--Mills Theory, in each case in flat spacetime viewed canonically, 
which realizes Configurational Relationalism but not Temporal Relationalism, i) 
\be
\mbox{the status of } \m {\cal S}^{\st\sr\si\sa\sll}_{\sR\se\sll} \m \mbox{ can be upgraded to } {\cal S}_{\sC\sC-\sR\si} 
\ee 
\n ii) The $\Shuffle$ arising from this action closes (by itself in these two examples), constituting moreover not just a $\bFlin$ but also a $\bGauge$.  

\m 

\n 3) For such as RPM or GR-as-Geometrodynamics, which implement both Temporal and Configurational Relationalism, the Relationalism-implementing trial action 
\be 
{\cal S}^{\st\sr\si\sa\sll}_{\sR\se\sll} \m \mbox{ is promoted to } \m  {\cal S}_{\sC\sC-\sR\si} \m , 
\ee
ii) Both cases' $\bShuffle$ self-closes and is confirmed to be of not only $\bFlin$ but $\bGauge$. 

\m 

\n RPM's $\Chronos = \scE$ also self-closes, whereas GR's $\Chronos = \scH$ requires GR's $\bGauge = \u{\bscM}$ as an integrability.  

\m 

\n The two furthermore mutually-close. 

\m 

\n The corresponding Best Matching is now confirmed to have the status 
\beq
{\cal S}_{\sC\sC\mbox{-}\sR\si} \:=
\lE_{\sbfg \, \in \, \nFrg} (\mbox{${\cal S}_{\sC\sC\mbox{-}\sR\si}$ built upon $\FrQ, \lFrg$})  \m ,
\label{E-Symbol-3-TR}
\eeq
where, 
\be 
\lE_{\sbfg \, \in \, \nFrg}  \es  \{\mbox{extremum over }  \lFrg\} \m {\cal S}_{\sC\sC-\sR\si}
\ee 
involves a suitable group action of $\lFrg$, and the whole construct has succeeded in getting past the TRi-Dirac Algorithm.  

\m 

\n iii) Finally, these theories' $\Chronos$ is rearranged to form a Machian emergent time of the conceptual form 
\beq
\ft^{\se\sm}_{\sC\sC\mbox{-}\sR\si}  \:=  \lE^{\prime}_{\sbig \, \in \, \nFrg} \m \ft^{\se\sm}_{\mbox{\scriptsize trial--}\nFrg}
\eeq
for 
$$
\ft^{\se\sm}_{\mbox{\scriptsize trial--}\nFrg}  \:=  \int \frac{||\d_{\sbig}\biQ||_{\sbiM}}{ \sqrt{2 \, W(\biQ)} }   \m , 
$$ 
and where $\mbox{\large E}^{\prime}$ is now likewise protected by $\bGauge$ closure. 

\m 

\n 3)'s full combined implementation of the first three facets of the Problem of Time is summarized in Fig \ref{TR-CR-CC-Summary}. 
We use here Article I's colour coding to keep track of single facet contributions and which combinations of facets enter each composite entity involve.

{            \begin{figure}[!ht]
\centering
\includegraphics[width=1\textwidth]{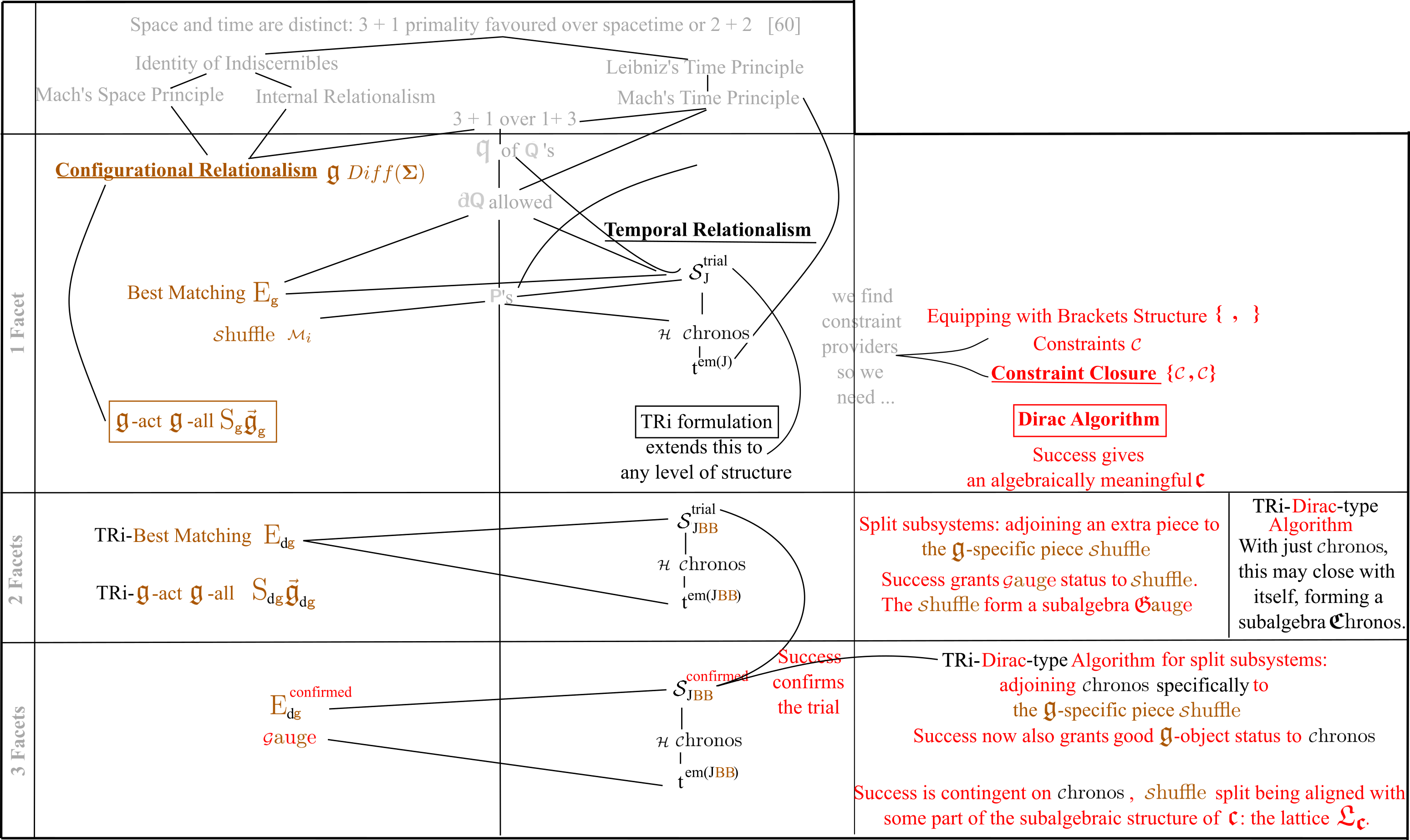} 
\caption[Text der im Bilderverzeichnis auftaucht]{        \footnotesize{A `technicolour guide' to how the various facet interferences of   
Constraint Closure           (red), 
Temporal Relationalism       (black), 
and  Configurational Relationalism (brown) fit together.
This figure can be viewed as an expansion of the leftmost portion of Fig III.6 concerning `in which order to address' Problem of Time facets. } }
\label{TR-CR-CC-Summary} \end{figure}          }

\m  

\n This position reached, each of Assignment of Observables and Spacetime Construction can be considered as separate extensions (Articles VIII and IX respectively).
See Article XIII for the overall resolution of the classical local Problem of Time's facet interference.  

\begin{appendices}

\section{Supporting Principles of Dynamics developments}

\subsection{The differential Hamiltonian}

TRi requires Hamiltonians to be replaced by {\it differential Hamiltonian} \cite{TRiPoD, MBook} change covectors,    
\be 
\ordial \cH[\bfQ, \bfP]  :=  \ordial \cJ[\bfQ, \ordial \bfQ] - \bfP \, \ordial \bfQ  \m .  
\ee
These remain rooted however on the same set of Hamiltonian variables, $(\bfQ, \bfP)$, these themselves being already-TRi.  
The differential Hamiltonian thus also lives on $\FrT^*(\FrQ)$.

\subsection{Passage to the anti-dRouthian}\label{anti-Routh}

{\bf Structure 1} The current Article also requires {\it passage to the anti-dRouthian} \cite{ABook}
\beq
\cA[\bar{\bfQ}, \bar{\bfP}, \d \bfc ]  \:=  \cL[\bar{\bfQ}, \d \bar{\bfQ}, \d \bfc ] - \bar{\bfP} \cdot \dot{\bar{\bfQ}}  \m .
\label{anti-Routhian} 
\eeq
\n{\bf Remark 1} Like passage to the dRouthian, this still involves treating the cyclic differentials as a separate package, 
albeit now under the diametrically opposite Legendre transformation.   
The anti-dRouthian completes the `Legendre square' whose other vertices are $\cL$, $\cH$ and $\cR$: Figure \ref{TRiPoD-Squares}.

\m 

\n{\bf Remark 2} Like passage to the Routhian, passage to the anti-(d)Routhian turns out to be a useful trick. 
A minor use is in Sec \ref{ARLM}, whereas the major use is in the next subsection.   
These uses are moreover {\sl specific foundational uses}, while Routhian tricks are useful in a {\sl concrete problem solving manner}.  

{            \begin{figure}[!ht]
\centering
\includegraphics[width=1\textwidth]{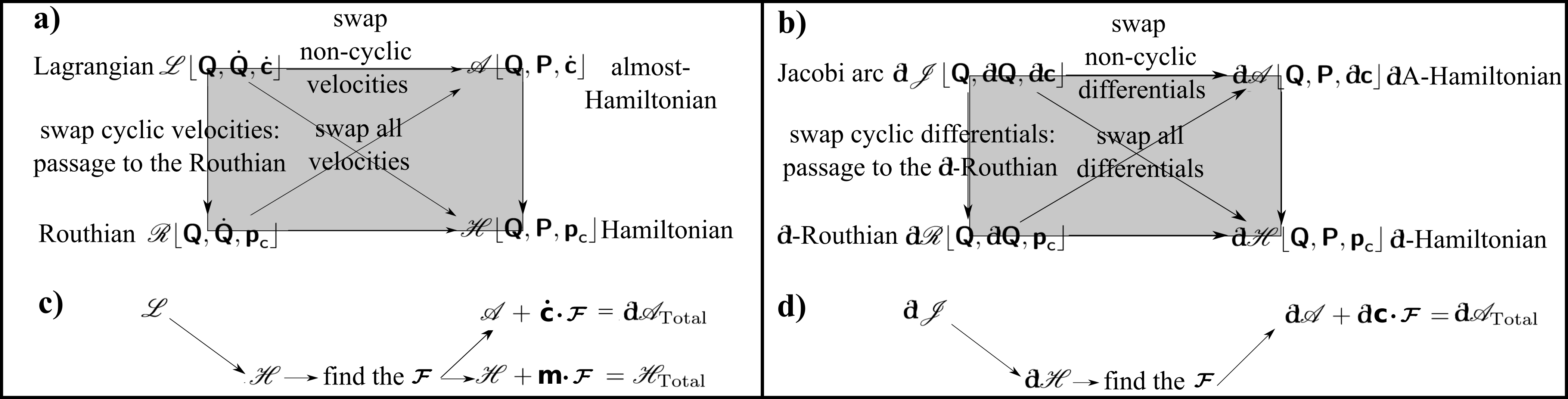}
\caption[Text der im Bilderverzeichnis auftaucht]{         \footnotesize{
a) Almost-Hamiltonian Legendre square.

\m 

\n b) then elevates this square to fully TRi form in terms of $\sordial$-Legendre transformations that dually switch momenta and changes. 
These are between change covectors: $\sordial\cJ$, $\sordial\cR$, $\sordial\cA$, $\sordial\cH$, the information-preserving extra terms now being subsystem Liouville forms, 
which were always change covectors. 
Routhians go to $\sordial$-Routhians;     
there is no need for `almost' in this case since Routhians are already allowed to contain velocities, and so already include almost-Routhians as a subset. 

\m 

\n 
c) and d) exhibit the choices by which the total Hamiltonian, A-Hamiltonian and $\sordial$A-Hamiltonian arise. 
The starred, primed and extended versions follow suit.} }
\label{TRiPoD-Squares}\end{figure}            }

\subsection{Further auxiliary spaces}\label{T-T*}

{\bf Structure 1} The Lagrangian and Hamiltonian variables respectively form the tangent bundle $\FrT(\FrQ)$ and cotangent bundle $\FrT^*(\FrQ)$ over $\FrQ$.  
From a geometrical perspective, the Legendre transformation for passage to the Hamiltonian is thus a map 
\be 
\FrT(\FrQ) \longrightarrow \FrT^*(\FrQ) \m .
\ee  
{\bf Definition 1} Let $\bFrC$ be the subconfiguration space of cyclic coordinates and $\check{\FrQ}$ be its complementary subconfiguration space in $\FrQ$.  

\m

\n{\bf Remark 1} The Routhian and the anti-Routhian tricks can now be seen to both come at a price. 
A first part of this price is geometrical: using these requires the following slightly more complicated {\it mixed cotangent--tangent bundles} over $\FrQ$. 

\m 

\n{\bf Structure 2}
\be 
\FrT(\check{\FrQ}) \times \FrT^*(\bFrC)
\ee 
is the bundle space for the Routhian \cite{ABook}, and 
\be 
\FrT^*(\check{\FrQ}) \times \FrT(\bFrC)
\ee 
is the bundle space for the anti-Routhian \cite{ABook}.  

\m

\n{\bf Remark 2} See Appendices \ref{Mor}-5 for the second and third parts of the price to pay

\subsection{Corresponding morphisms}\label{Mor}

{\bf Structure 1} The transformation theory for Hamiltonian variables is more subtle than that of the Lagrangian variables' $Point$ (already defined in Sec I.2.4).  
This in part reflects the involvement of  
\be
\bfP \cdot \dot{\bfQ}  
\label{P-dotQ}
\ee
due to its featuring in the conversion from $\cL$ to $\cH$. 

\m 

\n 1) Starting from $Point$, one can have the momenta follow suit so as to preserve (\ref{P-dotQ}) \cite{Lanczos}; 
these transformations indeed preserve $\cH$.

\m 

\n 2) Starting from $Point_t$, however, induces gyroscopic corrections to $\cH$ \cite{Lanczos}; this illustrates that $\cH$ itself can change form.

\m 

\n 3) More general transformations which mix the $\bfQ$ and the $\bfP$ are also possible.
These are however not as unrestrictedly           general functions of their $2 \, k$ arguments 
                      as $Point$'s transformations are as functions of their $k$      arguments.

\m 					  
					  
\n 3.i) The transformations which preserve the {\it Liouville 1-form} 
\beq
\bfP \cdot \ordial \bfQ  \m ,  
\label{Liouville}
\eeq
that is clearly associated with (\ref{P-dotQ}).
These can again be time-independent (termed {\it scleronomous}) 
or time-dependent in the sense of parametrization adjunction of $t$ to the $\bfQ$ (termed {\it rheonomous}). 
Again, the former preserve $H$ whereas the latter induce correction terms \cite{Lanczos}.  
These are often known as {\it contact transformations}, so we denote them by $Contact$ and $Contact_t$ respectively.  

\m 

\n 3.ii) More generally still, preserving the integral of (\ref{Liouville}) turns out to be useful for many purposes \cite{Lanczos}.
At the differential level, this corresponds to      (\ref{Liouville}) itself being preserved 
up to an additive complete differential $\d G$ for $G$ the {\it generating function}.
In this generality, one arrives at the {\it canonical transformations} alias {\it symplectomorphisms}, 
once again in the form a rhenonomous group with a scleronomous subgroup. 
We denote these by $Can_t$ and $Can$ respectively; see Fig \ref{PoD-Morphisms-Latt} for how this Sec's groups fit together to form a lattice of subgroups.  
%
{            \begin{figure}[!ht]
\centering
\includegraphics[width=0.27\textwidth]{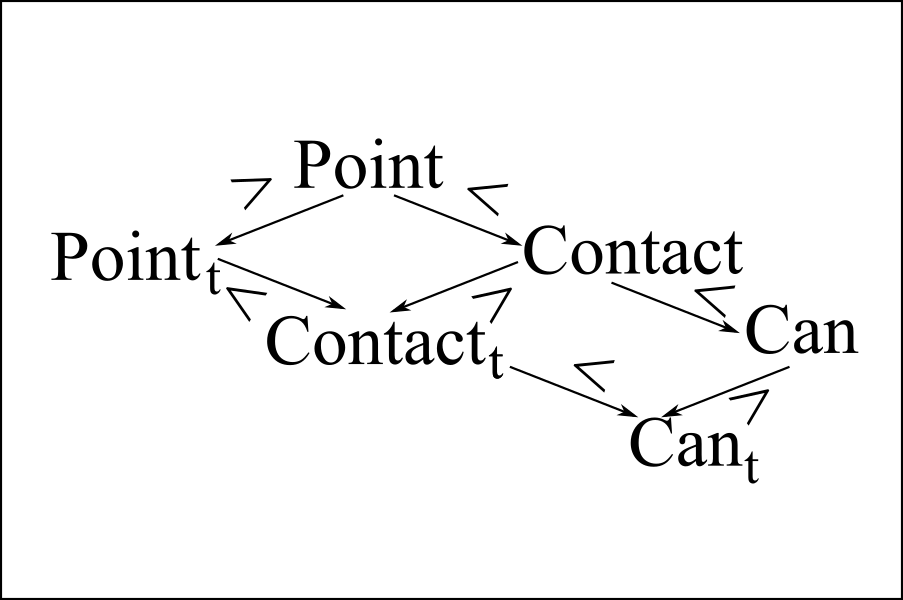}
\caption[Text der im Bilderverzeichnis auftaucht]{         \footnotesize{ 
\n Lattice structure of subgroups of Principles of Dynamics morphisms of varying generality.
}}
\label{PoD-Morphisms-Latt}\end{figure}            }

\m 

\n{\bf Remark 1} Whereas arbitrary canonical transformations do not permit explicit representation, infinitesimal ones do.  

\m 

\n{\bf Remark 2} Applying Stokes' Theorem to the integral of (\ref{Liouville}) reveals a more basic invariant: 
the bilinear antisymmetric {\it symplectic 2-form} \cite{Arnold}
\beq
\ordial \bfP \wedge \ordial \bfQ  \m . 
\label{Symplectic}
\eeq
We denote this by 
\be
\biomega \m \mbox{ with components } \m \omega_{\sfK\sfK^{\prime}}
\ee 
where the $\fK$ indices run over 1 to $2k$. 
This subsequently features in bracket structures (see e.g.\ two Section further down).

\m 

\n{\bf Remark 3} Concentrating on the $t$-independent case that is central to this Series of Articles, the morphisms for the Routhian formulation are 
\be 
Point(\check{\sFrQ}) \times Can(\bFrC)  \m .
\ee  
The latter piece is usually ignored due to the $\bic$ being absent and the $\bip^c$ being constant.

\m 

\n{\bf Structure 1} For the $t$-independent anti-Routhian formulation, the morphisms are 
\be 
Can(\check{\sFrQ}) \times Point(\bFrC) \m .
\ee   
These more complicated morphisms are the second price to pay in considering Routhian or anti-Routhian formulations.

\subsection{Peierls brackets}\label{Peierls}

\n{\bf Structure 1} A brackets structure can in fact already be associated with the Lagrangian tangent bundle formulation: 
the {\it Peierls bracket} \cite{Peierls, EMS02, DeWittBook}. 

\m 

\n{\bf Remark 1} This is more complicated than the Poisson bracket through involving Green's functions.  
Its explicit form is not required for this Series.  

\m 

\n{\bf Remark 2}  The third part of the price to pay if one uses a Routhian or anti-Routhian   
is that the mixed cotangent--tangent bundle nature of the variables requires in general {\it mixed Poisson--Peierls brackets}.

\subsection{(Anti-)Routhian analogue of the Legendre matrix}\label{ARLM}

The passage to the Hamiltonian is well-known to be affected by whether the Legendre matrix is invertible.

\m 

\n We now consider instead whether passage to the (anti-)Routhian is affected as well \cite{TRiPoD, ABook}.

\m 

\n{\bf Structure 1} The {\it Legendre matrix for the Routhian} is  
$$
\u{\u{\bslLambda}}  \:=  \frac{\pa^2 L}{\partional \dot{\u{\bfc}} \partional \dot{\u{\bfc}}}                 \m , 
$$ 
which is zero by (I.35), so this matrix is an relatively uninteresting albeit entirely obstructive object.

\m 

\n The corresponding expressions for acceleration are similarly entirely free of reference to the cyclic variables.  

\m 

\n On the other hand, the Legendre matrix for the anti-Routhian is 
$$
\u{\u{\bslLambda}}  \:=  \frac{\pa^2 L}{\partional \dot{\u{\bar{\bfQ}}} \partional \dot{\u{\bar{\bfQ}}}}  \m , 
$$ 
which is in general nontrivial. 

\m 

\n{\bf Remark 1} A theory of primary constraints can be based on this rather than on the usual larger Legendre matrix (I.69).

\m 

\n{\bf Remark 2} The smaller anti-Routhian trick is the observation that the acceleration of $\bar{\bfQ}$ is unaffected by the cyclic variables.
I.e.\ one can take (I.71) again with index $\fX$ in place of $\fA$ since the further terms involving the cyclic variables 
arising from the chain rule are annihilated by (I.35).

\subsection{$\lordial$A-Hamiltonians and phase spaces}\label{dA-Dir}

{\bf Structure 1} The Legendre matrix encoding the non-invertibility of the momentum-velocity relations is now supplanted by the 
{\it $\ordial^{-1}$-Legendre matrix} \cite{TRiPoD, ABook} change vector
\beq
\ordial^{-1} \u{\u{\bslOmega}}  \:=  \frac{\partional^2 \ordial \fJ}{\partional \, \ordial \u{\bfQ} \partional \, \ordial \u{\bfQ} } \m 
\left(
                          =  \frac{\partional \u{\bfP}}{\partional \, \ordial \u{\bfQ} } 
\right)  
\eeq
which encodes the non-invertibility of the momentum-change relations.

\m 

\n{\bf Structure 2} The TRi definition of primary constraint then follows 
in parallel to how the usual definition of primary constraint follows from the Legendre matrix, with secondary constraint remaining defined by exclusion. 

\m

\n{\bf Example 1} Dirac's argument that Reparametrization Invariance implies at least one primary constraint is now recast as Lemma I.5.  
The specific form of the primary constraint is, of course, $\Chronos$.  

\m 

\n{\bf Remark 1} The next idea in building a TRi version of Dirac's general treatment of constraints 
is to append constraints to one's incipient $\ordial$-Hamiltonian not with Lagrange multipliers -- which would break TRi -- but rather with cyclic differentials.
In this way, a $\ordial${\it A-Hamiltonian} is formed; 
the `A' here stands for `almost', though the $\ordial$A-Hamiltonian is also a particular case of $\ordial$-anti-Routhian. 

\m 

\n The $\ordial$A-Hamiltonian $\ordial A$ symbol aditionally has an extra minus sign relative to the $\ordial$-anti-Routhian $\ordial A$ symbol.
This originates from the definition of Hamiltonian involving an overall minus sign where the definitions of Routhian and anti-Routhian have none.  

\m 

\n Furthermore, in the current context, all the cyclic coordinates involved have auxiliary status and occur in $\lFrg$-correction combinations.
[In the event of a system possessing physical as well as auxiliary cyclic coordinates, one would use a `partial' rather than `complete' anti-Routhian.]  

\m 

\n{\bf Structure 3} The equations of motion are now $\ordial${\it A-Hamilton's equations}  \cite{TRiPoD, ABook} 
\beq
\ordial {\bfQ}  \es    \frac{\partional \ordial\cA}{\partional \, \bfP}   \mma
\ee
\be  
\ordial {\bfP}  \es  - \frac{\partional \ordial\cA}{\partional \, \bfQ}   \m , 
\label{dA-Ham-eqs}
\eeq
augmented by 
$$
\frac{\partional \, \ordial\cA}{\partional \, \ordial\bfc}  \es  0    \m .   
$$
\n{\bf Remark 2} Appendix \ref{ARLM}'s comment about using          the           anti-Routhian's own Legendre matrix carries over to 
                                                                    the $\ordial$-anti-Routhian,   
and thus also to the further identification of a subcase of this as the $\ordial$A-Hamiltonian.

\subsection{TRi-morphisms and brackets. ii)}

{\bf Remark 1} Suppose there are now cyclic differentials to be kept, or which arise from the TRi-Dirac Algorithm.
The corresponding morphisms are now a priori of the mixed type \c{ABook}
\be 
Can(\FrT^*(\FrQ)) \times Point(\lFrg)  \m .
\ee 
\n{\bf Remark 2} The brackets on these spaces are a priori of the mixed Poisson--Peierls type: Poisson as regards $\bfQ, \bfP$ and Peierls as regards $\ordial\bfg$. 

\m

\n{\bf Remark 3} (\ref{key}) implies that, as regards the constraints, 
\be 
Can(\FrT^*(\FrQ)) \times Point(\lFrg)
\ee 
reduces to just $Can(\FrT^*(\FrQ))$ and the mixed brackets reduce to just Poisson brackets on $\bfQ, \bfP$.   
{\sl The physical part of the $\ordial$A-Hamiltonian's incipient bracket is just a familiar Poisson bracket}.  
This good fortune follows from the $\ordial$A-Hamiltonian being a type of $\ordial$-anti-Routhian, 
alongside its non-Hamiltonian variables absenting themselves from the constraints due to the best-matched form of the action.  
 
\end{appendices}


\end{document}